\documentclass[reprint,amsmath,amssymb,aps,prd,floatfix]{revtex4-2}

\usepackage{graphicx}
\graphicspath{{./}}
\usepackage{dcolumn}
\usepackage{bm}
\usepackage{booktabs}
\usepackage{CJKutf8}
\usepackage{tikz,xcolor}
\usepackage{soul}
\sethlcolor{yellow}

\usepackage{hyperref}

\hypersetup{
  colorlinks=true,
  linkcolor=blue,
  urlcolor=blue,
  citecolor=blue,
  linktoc=all
}

\definecolor{lime}{HTML}{A6CE39}
\DeclareRobustCommand{\orcidicon}{%
  \begin{tikzpicture}
  \draw[lime, fill=lime] (0,0)
  circle [radius=0.16]
  node[white] {{\fontfamily{qag}\selectfont \tiny ID}};
  \draw[white, fill=white] (-0.0625,0.095)
  circle [radius=0.007];
  \end{tikzpicture}
  \hspace{-2mm}
}

\foreach \x in {A, ..., Z}{%
  \expandafter\xdef\csname orcid\x\endcsname{\noexpand\href{https://orcid.org/\csname orcidauthor\x\endcsname}{\noexpand\orcidicon}}
}

\begin{document}
\begin{CJK*}{UTF8}{gbsn}

\title{Repeated Gravitational-Wave Lensing from Binary Black Holes on Eccentric and Precessing Orbits around Spinning Supermassive Black Holes}
\author{Jie Wu (吴洁)\orcidA{}$^{1,2}$ }
\author{Mengfei Sun (孙孟飞)\orcidB{}$^{1,2}$ }
\author{Jin Li (李瑾)\orcidC{}$^{1,2,3}$ }
\email{cqujinli1983@cqu.edu.cn}
\author{Zhoujian Cao (曹周键)\orcidD{}$^{4,5,6}$}
\email{zjcao@bnu.edu.cn}
\affiliation{$^{1}$College of Physics, Chongqing University, Chongqing 401331, China}
\affiliation{$^{2}$Department of Physics and Chongqing Key Laboratory for Strongly Coupled Physics, Chongqing University, Chongqing 401331, China}
\affiliation{$^{3}$Institute of Advanced Interdisciplinary Studies, Chongqing University, Chongqing 401331, China}
\affiliation{$^{4}$Department of Astronomy, Beijing Normal University, Beijing 100875, China}
\affiliation{$^{5}$Institute for Frontiers in Astronomy and Astrophysics, Beijing Normal University, Beijing 102206, China}
\affiliation{$^{6}$School of Fundamental Physics and Mathematical Sciences, Hangzhou Institute for Advanced Study, UCAS, Hangzhou 310024, China}
\begin{abstract}
Gravitational waves (GWs) from a stellar-mass binary black hole (BBH) orbiting a supermassive black hole (SMBH) can experience repeated gravitational lensing as the BBH passes behind the SMBH. Our previous studies of orbit-induced GW modulation motivate this investigation of repeated lensing in evolving hierarchical triple systems. We investigate how eccentricity, inclination, and relativistic precession of the outer orbit affect successive lensing events and parameter inference. The outer orbit follows post-Newtonian dynamics, and the inspiral waveform includes moving-source effects and de Sitter precession. An event-local point-mass lensing model tracks the changing source--lens alignment and GW frequency. For a one-year LISA observation, we examine lensing-event identification, Fisher-matrix estimates of parameter uncertainties, and waveform residuals when lensing is omitted. Our results show that outer-orbit evolution changes the timing and amplification of successive lensing events and can bring initially unfavorable configurations into the lensing region. A fixed circular, nonprecessing approximation misses all lensing events for a substantial fraction of the selected lensed systems. The median fractional reductions in uncertainty are largest for the outer-orbit inclination and SMBH spin. For some systems above the adopted detection threshold, lensing-induced waveform differences persist after local adjustments to the parameters of the unlensed model. Repeated lensing can thus provide complementary constraints on SMBH properties and outer-orbit geometry in LISA observations of hierarchical BBHs.
\end{abstract}
\maketitle
\end{CJK*}

\section{Introduction}

The first direct detection of gravitational waves (GWs) from a merging binary black hole (BBH) provided a new way to observe compact binaries~\cite{Abbott2016GW150914}. The Laser Interferometer Space Antenna (LISA) is expected to extend GW observations to millihertz frequencies, where it can follow stellar-mass BBHs during their early inspiral~\cite{Colpi2024LISA,Sesana2016Multiband}. Suitable signals can remain in band for much of an observation, allowing slow frequency evolution to be tracked and weak perturbations to accumulate~\cite{Toubiana2020StellarLISA,Toubiana2021Environmental}. One configuration of interest is a BBH orbiting a supermassive black hole (SMBH), whose properties are encoded in the outer orbital motion.

The outer orbit modulates the observed GW signal through kinematic and relativistic effects. Over a sufficiently long observation, the line-of-sight acceleration of the BBH center of mass can produce a measurable phase shift~\cite{Inayoshi2017Acceleration,Tamanini2020Acceleration}. More generally, Lorentz transformations relate the waveform of a moving source to its rest-frame waveform~\cite{He2022Moving}. As the outer-orbit velocity varies, Doppler modulation traces the source motion, while transverse velocity provides information beyond the leading line-of-sight contribution~\cite{YuChen2021SMBH,KuntzLeyde2023Transverse}. Beyond these kinematic effects, de Sitter (dS) precession changes the orientation of the inner binary and hence its waveform~\cite{YuChen2021SMBH}. The SMBH spin further changes the outer orbit through frame dragging~\cite{Fang2019Spin}. Eccentricity makes the modulation depend on orbital phase, while relativistic periapsis precession changes the source geometry between successive outer revolutions~\cite{Laeuger2024Eccentric}. Outer-orbit dynamics therefore governs both the waveform modulation and the evolving source--lens--observer geometry.

When the BBH passes behind the SMBH with a sufficiently small projected source--lens separation, gravitational lensing of the emitted GWs can occur~\cite{DOr2020Repeated,Yu2021Detecting}. The source--lens alignment changes continuously as the outer orbit evolves, and favorable alignments may recur during a long observation. Wave effects become important when the GW period is comparable to the characteristic lensing time delay~\cite{Nakamura1998Point,NakamuraDeguchi1999Wave}. For a point-mass lens, a frequency-dependent complex amplification factor modifies both the amplitude and phase of the signal~\cite{Takahashi2003Wave}. For an inspiraling BBH, the evolving GW frequency samples different parts of this response while the outer orbit simultaneously changes the source--lens alignment~\cite{DOr2020Repeated,Takahashi2003Wave}. At sufficiently high frequencies, the wave-optics amplification approaches the geometric-optics limit, with images of different magnifications, time delays, and phases~\cite{Leung2025WaveOptics}. Successive passages behind the SMBH can therefore produce distinct lensing signatures.

Previous studies have examined GW lensing and its implications for parameter inference. Cao et al. studied its effects on signal recovery and parameter estimation with ground-based detectors~\cite{Cao2014Lensing}. For long-duration space-based observations, repeated lensing in SMBH--BBH triples has been investigated for circular inner and outer orbits~\cite{DOr2020Repeated}. Yu \textit{et al.} connected the magnification to triple waveforms with Doppler and dS precession and examined parameter inference for circular configurations~\cite{Yu2021Detecting}. Pijnenburg \textit{et al.} investigated wave-optics lensing and polarization structure in LISA-band triples, going beyond a single scalar amplification factor~\cite{Pijnenburg2024Wave}. Li \textit{et al.} showed that Lense--Thirring (LT) precession can modulate repeated lensing of continuous-wave sources on nearly circular outer orbits~\cite{Li2026LT}. Population-level work has also estimated self-lensing probabilities for compact binaries in galactic nuclei, including active galactic nuclei~\cite{Ubach2025SelfLensing}. Other strong-field calculations examine lensing, aberration, and polarization effects near massive black holes in different regimes~\cite{Oancea2024SpinHall,Santos2025BEMRI,Huang2026SelfLensing}. Eccentric outer orbits and periapsis precession have also been studied in triple waveform modeling and parameter inference~\cite{Laeuger2024Eccentric}. The repeated-lensing studies cited above adopt circular or nearly circular outer orbits. The cited eccentric-orbit study instead examines dynamical waveform modulation without tracking the lensing-event sequence. A complementary question is how eccentricity, inclination, and relativistic precession affect successive lensing events and parameter inference for an inspiraling stellar-mass BBH.

Our previous work used orbit-induced GW modulation from stellar-mass BBHs to probe globular-cluster properties~\cite{Wu2025Globular}. We subsequently studied BBHs orbiting Kerr SMBHs, relating relativistic orbital modulation to constraints on the central black hole and outer orbit~\cite{Wu2026Kerr}. An extension to AGN disks incorporated gaseous dynamical friction and examined its implications for environmental inference and waveform-model errors~\cite{Wu2026AGN}. These studies motivate us to investigate repeated GW lensing as the source--lens alignment changes along the outer orbit. We consider an inspiraling stellar-mass BBH on an eccentric, inclined, and precessing orbit around a Kerr SMBH. We compare the evolving model with a simplified circular outer-orbit baseline to assess how orbital evolution changes event occurrence and amplification. The analysis also examines how lensing affects LISA parameter uncertainties and what residual waveform structure remains after local parameter shifts in an unlensed model.

The paper is organized as follows. We describe the hierarchical triple and the GW signal in Sec.~\ref{sec:triple_signal}. Section~\ref{sec:lensing_waveform} presents the lensing model and waveform construction. We examine repeated lensing in evolving outer orbits in Sec.~\ref{sec:repeated_lensing}. Section~\ref{sec:inference} discusses the implications for parameter inference and the waveform residuals that remain when lensing is neglected. Finally, we summarize the results and their scope in Sec.~\ref{sec:conclusion}. Appendix~\ref{app:eccentric_lensing_probability} derives the approximate lensing probability for an eccentric outer orbit and specifies the adopted orientation convention. Throughout this paper, we use geometrized units with $c=G=1$.

\section{Hierarchical triple and gravitational-wave signal}
\label{sec:triple_signal}

\subsection{Outer-orbit geometry and dynamics}
\label{subsec:triple_orbit}

\begin{figure*}[t]
\centering
\includegraphics[width=0.96\textwidth]{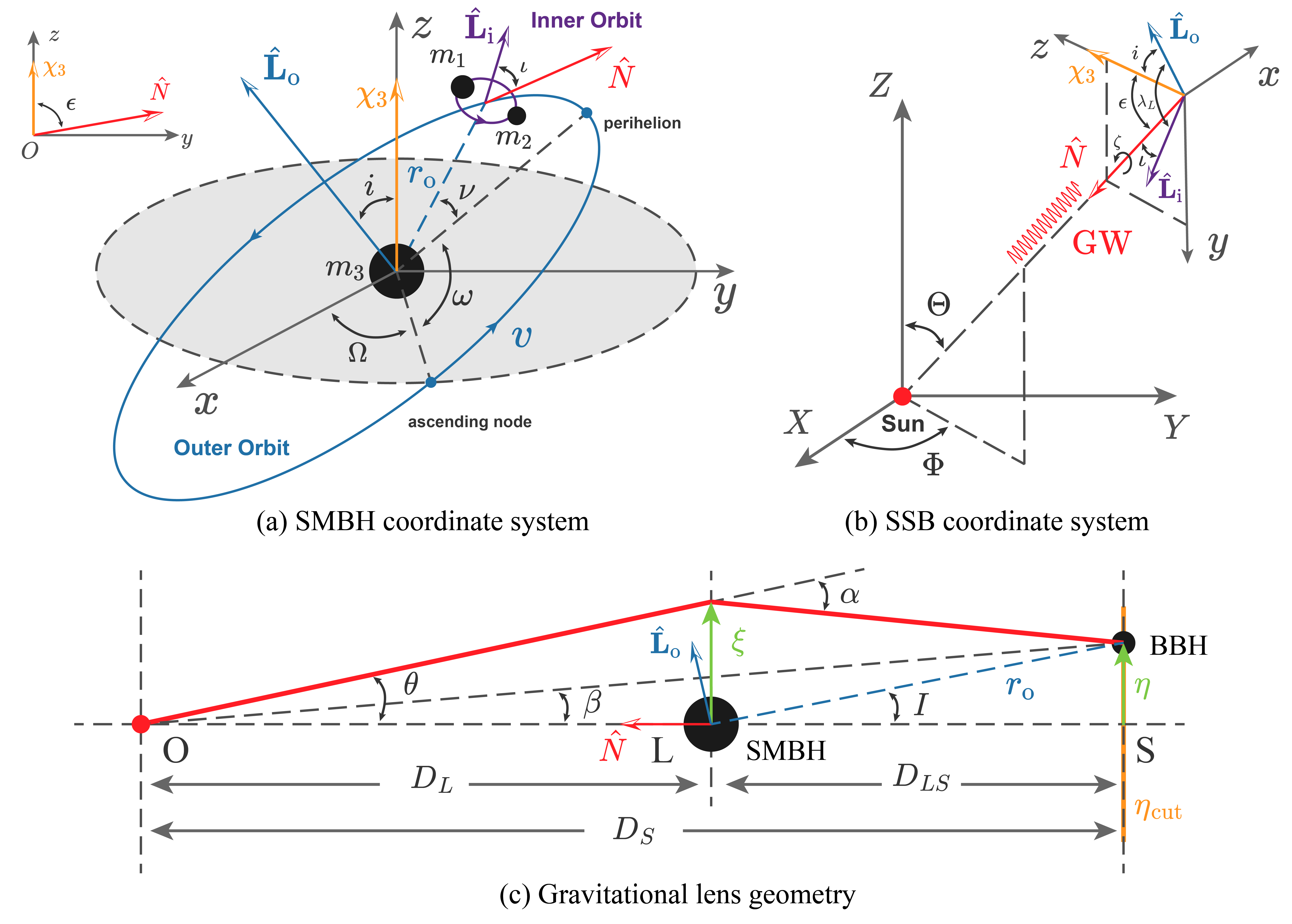}
\caption{Schematic of the SMBH--BBH hierarchical triple and the coordinate systems used in this work. In (a) and (b), we show the SMBH-centered $(x,y,z)$ and Solar System barycenter (SSB) $(X,Y,Z)$ frames, respectively. The $y$ axis lies in the plane containing $z$ and the GW propagation direction $\hat{\mathbf N}$, while $\hat{\mathbf x}=\hat{\mathbf y}\times\hat{\mathbf z}$. (c) shows the source--lens--observer geometry, with O, L, and S denoting the observer, lens, and source. The red lines indicate GW propagation paths. All elements are schematic and not to scale.}
\label{fig:lens_geometry}
\end{figure*}

We consider the SMBH--BBH hierarchical triple shown in Fig.~\ref{fig:lens_geometry}. A stellar-mass BBH with component masses $m_1$ and $m_2$ forms the inner binary, while its center of mass orbits an SMBH of mass $m_3$. We use subscripts ${\rm i}$ and ${\rm o}$ for inner- and outer-orbit parameters, respectively. Indices 1 and 2 identify the stellar-mass black holes, whereas 3 identifies the SMBH. The system lies in the hierarchical regime $m_3\gg m=m_1+m_2$ and $a_{\rm o}\gg a_{\rm i}$, with well-separated inner and outer orbital scales. The model is restricted to dynamically stable hierarchical triples~\cite{EggletonKiseleva1995}. Lidov--Kozai oscillations may occur in such systems~\cite{Prodan2015Secular}, but we neglect the associated secular evolution of the inner binary over the one-year observation, following Refs.~\cite{Wu2026Kerr,Wu2026AGN}.

In the SMBH-centered frame of Fig.~\ref{fig:lens_geometry}(a), we choose the SMBH spin direction as the $z$ axis and denote its dimensionless magnitude by $\chi_3$. The outer orbit is specified by the semimajor axis $a_{\rm o}$, eccentricity $e_{\rm o}$, inclination $i$, longitude of the ascending node $\Omega$, argument of periapsis $\omega$, and true anomaly $\nu$. The inclination $i$ is the angle between the outer orbital angular momentum and the SMBH spin axis, with $\cos i=\hat{\mathbf L}_{\rm o}\cdot\hat{\mathbf z}$. The GW propagation direction $\hat{\mathbf N}$ is shown in Fig.~\ref{fig:lens_geometry}(b). In the $yz$ plane, $\epsilon$ is the angle between $\hat{\mathbf N}$ and the SMBH spin axis. The angle $\zeta$ fixes the remaining rotation about $\hat{\mathbf N}$, while $(\Phi,\Theta)$ specifies the source sky position in the SSB frame. The four angles $\epsilon$, $\zeta$, $\Phi$, and $\Theta$ determine the transformation of the outer-orbit position and velocity from the SMBH-centered frame to the SSB frame~\cite{Wu2026Kerr}.

We evolve the outer orbit numerically by solving the first-order system
\begin{equation}
\dot{\mathbf r}_{\rm o}=\mathbf v_{\rm o},\qquad
\dot{\mathbf v}_{\rm o}=\mathbf a_{\rm o}.
\end{equation}
We include relativistic effects associated with the central SMBH through a post-Newtonian (PN) expansion of the acceleration~\cite{Blanchet2014PN},
\begin{equation}
\mathbf a_{\rm o}=\mathbf a_{\rm N}+\mathbf a_{\rm 1PN}+\mathbf a_{\rm SO}+\mathbf a_{\rm 2PN}+\mathbf a_{\rm SS}+\mathbf a_{\rm QM}.
\end{equation}
The term $\mathbf a_{\rm N}$ denotes the Newtonian contribution, while $\mathbf a_{\rm 1PN}$ and $\mathbf a_{\rm 2PN}$ are the conservative nonspinning PN corrections. The terms $\mathbf a_{\rm SO}$, $\mathbf a_{\rm SS}$, and $\mathbf a_{\rm QM}$ account for spin-orbit, spin-spin, and quadrupole--monopole couplings, respectively~\cite{Kidder1995Spin,GergelyKeresztes2003}. Outer-orbit radiation reaction is neglected over the one-year observation. Relativistic periapsis precession and spin-induced evolution of the orbital plane affect both waveform modulation and lensing geometry.
The numerical evolution yields the time-dependent BBH center-of-mass position $\mathbf r_{\rm o}$ and velocity $\mathbf v_{\rm o}$. After transformation to the SSB frame, they determine the waveform modulation discussed below and the evolving source--SMBH--observer geometry used in the lensing analysis.

\subsection{Gravitational-wave signal}
\label{subsec:gw_signal}

LISA can observe stellar-mass BBHs during their early inspiral, which can be described with PN theory~\cite{Blanchet2014PN}. In the BBH rest frame, the time-domain inspiral waveform of a quasicircular inner binary with aligned component spins can be written as
\begin{equation}
h_0=\mathcal{A}\,\mathrm{e}^{i\phi_{\rm PN}},
\end{equation}
with amplitude
\begin{equation}
\mathcal{A}=\frac{4\mathcal{M}_c^{5/3}}{D_L}(\pi f)^{2/3}.
\end{equation}
The chirp mass is $\mathcal{M}_c=(m_1m_2)^{3/5}/(m_1+m_2)^{1/5}$, $D_L$ is the luminosity distance, and $2\pi f=\dot\phi_{\rm PN}$. We use nonspinning phase evolution through 3.5PN order, supplemented by linear spin-orbit and quadratic spin-spin contributions from the aligned components $\chi_{1z}$ and $\chi_{2z}$. This is denoted the 3.5PN+SO+SS waveform~\cite{Wu2024PNComparison}, with explicit PN expansions given in Refs.~\cite{Blanchet2014PN,ChoPortoYang2022Spin}. The outer orbit produces Doppler modulation, which can yield antichirping intervals during which the observed GW frequency temporarily decreases~\cite{Toubiana2021Environmental}. A global stationary-phase approximation can then become inadequate~\cite{Sberna2022Antichirping}, so we construct the signal in the time domain and Fourier transform it as needed.

As the BBH moves in the gravitational field of the SMBH, its inner orbital angular momentum $\hat{\mathbf L}_{\rm i}$ undergoes dS precession~\cite{YuChen2021SMBH,Laeuger2024Eccentric}. We evolve its direction using the instantaneous outer-orbit position and velocity,
\begin{equation}
\frac{\mathrm{d}\hat{\mathbf L}_{\rm i}}{\mathrm{d}t}=\boldsymbol{\Omega}_{\rm dS}\times\hat{\mathbf L}_{\rm i}.
\end{equation}
The precession vector is~\cite{BarkerOConnell1975}
\begin{equation}
\boldsymbol{\Omega}_{\rm dS}=\frac{3}{2}\frac{m_3+\mu_{\rm o}/3}{r_{\rm o}^3}\left(\mathbf r_{\rm o}\times\mathbf v_{\rm o}\right).
\end{equation}
In this expression, $\mu_{\rm o}=m_3m/(m_3+m)$ and $r_{\rm o}=|\mathbf r_{\rm o}|$. For a Keplerian outer orbit, orbital averaging recovers the standard orbit-averaged dS precession rate used in Ref.~\cite{YuChen2021SMBH} in the hierarchical mass limit. We evaluate the instantaneous expression along the numerically evolved eccentric and precessing outer orbit. The changing direction of $\hat{\mathbf L}_{\rm i}$ also alters the inclination $\iota$ relative to the GW propagation direction $\hat{\mathbf N}$, where $\cos\iota=\hat{\mathbf L}_{\rm i}\cdot\hat{\mathbf N}$. The polarization amplitudes are therefore
\begin{equation}
A_+=\frac{1+(\hat{\mathbf L}_{\rm i}\cdot\hat{\mathbf N})^2}{2},\qquad A_\times=\hat{\mathbf L}_{\rm i}\cdot\hat{\mathbf N}.
\end{equation}
This precession also changes the polarization geometry and contributes the Thomas phase $\phi_T$~\cite{Apostolatos1994Precession,YuChen2021SMBH}.

LISA responds to the two GW polarizations through the antenna response functions $F_+$ and $F_\times$, which depend on the apparent sky position and polarization basis~\cite{Cutler1998LISA,Liang2019Response}. We include the frequency-dependent finite-arm response for an arm length of $2.5\times10^9\,{\rm m}$~\cite{CornishLarson2001Response,Robson2019LISA}. The angle $\Psi$ specifies the polarization basis entering $F_+$ and $F_\times$. The antenna response functions combine with $A_+$ and $A_\times$ to give the complex response $A_+F_+-iA_\times F_\times$. Separately, LISA's orbital motion produces the annual Doppler phase~\cite{Cutler1998LISA}
\begin{equation}
\phi_D=2\pi f\,{\rm A.U.}\times\sin\Theta\cos\left(2\pi t/{\rm yr}-\Phi\right).
\end{equation}
This phase arises from LISA's heliocentric orbital motion projected along the source direction.

The motion of the BBH center of mass around the SMBH also modifies the received waveform, which we calculate using a moving-source transformation~\cite{He2022Moving,He2026Accelerating}. Aberration maps the apparent direction $(\Theta,\Phi)$ to $(\Theta',\Phi')$, while the Lorentz boost rotates the polarization basis from $\Psi$ to $\Psi'$. The transformed direction and polarization basis enter the antenna response functions $F_+$ and $F_\times$. In the complex-strain representation, the boost contributes $\mathrm{e}^{-2i\lambda}$, where $\lambda$ depends on the propagation direction and source velocity. Under constant-velocity motion, the constant phase $\lambda$ can be absorbed into the initial phase, whereas a varying outer-orbit velocity requires retaining its time dependence. The transformation also relates the source-frame coordinate time $t_{\rm S}$ to the detector-frame coordinate time $t_{\rm D}$~\cite{He2022Moving,Wu2025Kick},
\begin{equation}
\frac{\mathrm{d}t_{\rm S}}{\mathrm{d}t_{\rm D}}=\frac{1}{\gamma(1-\mathbf v\cdot\hat{\mathbf N})},\qquad \gamma=\frac{1}{\sqrt{1-v^2}}.
\end{equation}
The projection $\mathbf v\cdot\hat{\mathbf N}$ gives the longitudinal Doppler contribution, whereas $\gamma$ accounts for transverse-Doppler time dilation~\cite{Wu2026AGN}. The changing velocity thus determines which stage of the rest-frame inspiral corresponds to a given detector time.

To isolate the static gravitational-redshift contribution, we use the time component of the Kerr metric in Boyer--Lindquist coordinates~\cite{BoyerLindquist1967},
\begin{equation}
g_{tt}=-\left(1-\frac{2m_3r}{\Sigma}\right),\qquad \Sigma=r^2+(\chi_3m_3)^2\cos^2\theta.
\end{equation}
At the instantaneous BBH position, $r=r_{\rm o}$ and $\cos\theta=\hat{\mathbf r}_{\rm o}\cdot\hat{\mathbf z}$. We denote the proper time along the BBH center-of-mass worldline by $\tau_{\rm S}$, with $\mathrm{d}\tau_{\rm S}^2=-\mathrm{d}s^2$. The static time-dilation factor at that position is approximated by $\sqrt{-g_{tt}}$,
\begin{equation}
\frac{\mathrm{d}\tau_{\rm S}}{\mathrm{d}t_{\rm S}}=\sqrt{-g_{tt}}=\sqrt{1-\frac{2m_3r_{\rm o}}{r_{\rm o}^2+(\chi_3m_3)^2(\hat{\mathbf r}_{\rm o}\cdot\hat{\mathbf z})^2}}.
\end{equation}
Velocity-dependent time dilation is included separately through $\gamma$. The time mapping used in the waveform is
\begin{equation}
\frac{\mathrm{d}\tau_{\rm S}}{\mathrm{d}t_{\rm D}}=\frac{\mathrm{d}\tau_{\rm S}}{\mathrm{d}t_{\rm S}}\frac{\mathrm{d}t_{\rm S}}{\mathrm{d}t_{\rm D}}.
\end{equation}
Integrating this relation gives the source proper time corresponding to each detector time. Further details of the time transformation are given in Ref.~\cite{Wu2026AGN}. The resulting detector-frame signal is
\begin{equation}
h(t_{\rm D})=\left(A_+F_+-iA_\times F_\times\right)h_0[\tau_{\rm S}(t_{\rm D})]\,\mathrm{e}^{i(\phi_D-2\phi_T-2\lambda)}.
\end{equation}
The amplitudes $A_+$ and $A_\times$ encode the inner-orbit orientation. The response functions $F_+$ and $F_\times$ carry the apparent sky-position and polarization-basis dependence, including changes induced by aberration. The intrinsic carrier $h_0$ is evaluated at the mapped proper time. This detector-frame signal provides the unlensed baseline for the population analysis and the lensing construction in Sec.~\ref{sec:lensing_waveform}.

\subsection{Source population and sampling}

We construct a population of SMBH--BBH hierarchical triples to study repeated lensing across a range of configurations. The sampled quantities include the inner-binary, SMBH, and outer-orbit parameters, as well as the orientation angles and phases entering the waveform. Their distributions are summarized in Table~\ref{tab:population}, with the outer semimajor axis expressed in units of the SMBH Schwarzschild radius $R_s=2m_3$.

\begin{table}[h]
\caption{Parameter distributions used in this work. The stellar-mass BBH population choices are informed by Ref.~\cite{Babak2023Stochastic}. $U[a,b]$ denotes a uniform distribution over $[a,b]$.}
\label{tab:population}
\begin{tabular*}{\columnwidth}{@{\extracolsep{\fill}}ll@{}}
\toprule
Parameter & Distribution \\
\midrule
$m\,[\mathrm{M}_\odot]$ & $U[20,230]$ \\
$q_{\rm i}$ & $U[0.3,1]$ \\
$\chi_{1z},\chi_{2z}$ & $U[-0.9,0.9]$ \\
$\log_{10}(z)$ & $U[-2.17,-0.04]$ \\
$\tau\,[{\rm yr}]$ & $U[1,100]/(1+z)$ \\
$\Phi,\phi_0,\zeta,\Omega,\omega,\nu_0\,[{\rm rad}]$ & $U[0,2\pi]$ \\
$\Theta,\epsilon,\lambda_L,i\,[{\rm rad}]$ & $\arccos\!\left(U[-1,1]\right)$ \\
$\log_{10}(m_3/\mathrm{M}_\odot)$ & $U[5,9]$ \\
$\log_{10}(a_{\rm o}/R_s)$ & $U[1.47,3.47]$ \\
$\chi_3,e_{\rm o}$ & $U[0,0.95]$ \\
\bottomrule
\end{tabular*}
\end{table}

We derive the remaining waveform inputs from these parameters. The inner-BBH mass ratio is $q_{\rm i}=m_2/m_1$, and the component masses and chirp mass $\mathcal M_c$ are determined from $m$ and $q_{\rm i}$. With $\tau$ denoting the time to coalescence, we set the initial GW frequency using $\mathcal M_{c,z}=(1+z)\mathcal M_c$,
\begin{equation}
f_0=\frac{1}{8\pi}\left(\frac{1}{5}\mathcal M_{c,z}^{5/3}\tau\right)^{-3/8}.
\end{equation}

Cosmological redshift enters the detector-frame BBH waveform through the component masses \mbox{$m_{1,2,z}=(1+z)m_{1,2}$}. The luminosity distance is obtained from the sampled redshift $z$, assuming a flat cosmology with a cosmological constant and cold dark matter,
\begin{equation}
D_L=\frac{1+z}{H_0}\int_0^z\frac{\mathrm{d}z'}{\sqrt{\Omega_m(1+z')^3+\Omega_\Lambda}}.
\end{equation}
We adopt $H_0=67.36\,{\rm km\,s^{-1}\,Mpc^{-1}}$, $\Omega_m=0.315$, and $\Omega_\Lambda=0.685$ from the Planck 2018 results~\cite{Planck2020Cosmology}.

The outer-orbit elements and coordinate angles follow the conventions illustrated in Fig.~\ref{fig:lens_geometry}. The angle $\lambda_L$ is the initial angle between the inner and outer orbital angular momenta, with $\cos\lambda_L=\hat{\mathbf L}_{\rm i}\cdot\hat{\mathbf L}_{\rm o}$. The angle $\nu_0$ is the initial true anomaly of the outer orbit, while $\phi_0$ is the initial phase of the inner-BBH waveform.

For each sampled system, we generate a one-year LISA signal using the waveform model above. Waveforms are evaluated within the validity range of the adopted inspiral model. We quantify the signal strength with the standard noise-weighted inner product~\cite{Cutler1998LISA}
\begin{equation}
(a|b)=4\,\mathrm{Re}\int_0^\infty\frac{\tilde a^*(f)\tilde b(f)}{S_n(f)}\,\mathrm{d}f.
\end{equation}
The functions $\tilde a(f)$ and $\tilde b(f)$ are the Fourier transforms of the time-domain signals $a(t)$ and $b(t)$. The one-sided LISA noise power spectral density $S_n(f)$ includes instrumental noise~\cite{Robson2019LISA} and the Galactic compact-binary confusion foreground~\cite{WuLi2023Confusion}. The signal-to-noise ratio (SNR) is evaluated from the unlensed signal as
\begin{equation}
\mathrm{SNR}=\sqrt{(h|h)}.
\end{equation}

We exclude configurations that fail the adopted hierarchical-stability criterion~\cite{EggletonKiseleva1995}. The stable candidates are further preselected using the approximate lensing probability in Eq.~\eqref{eq:eccentric_lensing_opportunity}, with $P_{\rm L}>10^{-3}$, and a Keplerian scan of the source--lens geometry. We also include systems that may enter the lensing region during the observation due to LT precession. We randomly select systems without replacement from the preselected population for the subsequent analysis. To restrict the sample to the adopted PN regime, we define the periapsis expansion parameter $\varepsilon_{\rm PN}\equiv m_3/[a_{\rm o}(1-e_{\rm o})]$ and retain systems satisfying $\varepsilon_{\rm PN}\leq0.1$. Systems without lensing events along the numerically evolved PN orbit are then removed. The remaining lensed systems with $\mathrm{SNR}\geq8$ are selected for the repeated-lensing and parameter-inference analyses~\cite{Buscicchio2025StellarMass}. The lensing geometry, event criterion, and waveform modification are described in Sec.~\ref{sec:lensing_waveform}.

\section{Gravitational lensing and waveform construction}
\label{sec:lensing_waveform}

\subsection{Point-mass lensing geometry and amplification}
\label{subsec:point_lens}

Section~\ref{sec:triple_signal} gives the unlensed detector-frame signal, including the effects of the outer orbit. When the BBH passes behind the SMBH and approaches the optical axis connecting the lens and observer, its GWs can be lensed. These alignments may recur as the BBH completes successive outer orbits~\cite{DOr2020Repeated}. We adopt a point-mass lens model to describe the leading contribution of the SMBH mass to standard lensing in the weak-deflection approximation. This model provides a common description of wave-optics amplification and its geometric-optics limit. More compact outer orbits around massive SMBHs may also produce retrolensing signatures~\cite{Yu2021Detecting}.

Using the geometry shown in Fig.~\ref{fig:lens_geometry}(c), the line-of-sight separation between the lens and source is
\begin{equation}
D_{\rm LS}(t)=-\mathbf r_{\rm o}(t)\cdot\hat{\mathbf N}.
\end{equation}
Positive $D_{\rm LS}$ places the BBH behind the SMBH. Its transverse offset from the optical axis is
\begin{equation}
u(t)=\left|\mathbf r_{\rm o}(t)-[\mathbf r_{\rm o}(t)\cdot\hat{\mathbf N}]\hat{\mathbf N}\right|.
\end{equation}
The SMBH--BBH separation is negligible compared with their distance from the observer. In this near-source limit, the Einstein radius is
\begin{equation}
R_E(t)=\sqrt{4m_3D_{\rm LS}(t)}.
\end{equation}
We express the projected source position in units of this radius as
\begin{equation}
\eta(t)=\frac{u(t)}{R_E(t)}.
\end{equation}

Since smaller $\eta$ corresponds to closer alignment with the optical axis, we identify a standard lensing event as a contiguous time interval satisfying
\begin{equation}
D_{\rm LS}(t)>0,\qquad \eta(t)<\eta_{\rm cut}.
\end{equation}
We adopt $\eta_{\rm cut}=1.27$, which approximately corresponds to a secondary-image magnification $|\mu_-|\gtrsim0.1$ for a point-mass lens~\cite{Yu2021Detecting}.

In the LISA frequency band, wave effects can be important when the GW period is comparable to the image time delay~\cite{Nakamura1998Point}. We define $M_L=(1+z)m_3$ as the redshifted lens mass and $w=8\pi M_Lf$ as the dimensionless frequency. The point-mass wave-optics amplification takes the form~\cite{Takahashi2003Wave}
\begin{equation}
\begin{aligned}
F_{\rm wave}={}&\exp\!\left[\frac{\pi w}{4}
+\frac{iw}{2}\left(\ln\frac{w}{2}-2\phi_m\right)\right]\\
&\times\Gamma\!\left(1-\frac{iw}{2}\right)
{}_1F_1\!\left(\frac{iw}{2},1;\frac{iw\eta^2}{2}\right),
\end{aligned}
\label{eq:lens_wave}
\end{equation}
where $\phi_m=(x_m-\eta)^2/2-\ln x_m$, with $x_m=(\eta+\sqrt{\eta^2+4})/2$. The functions $\Gamma$ and ${}_1F_1$ denote the gamma and confluent hypergeometric functions, respectively. As $\eta(t)$ varies along the outer orbit, this complex factor modifies both the GW amplitude and phase.

\begin{figure}[!ht]
\centering
\includegraphics[width=0.96\columnwidth]{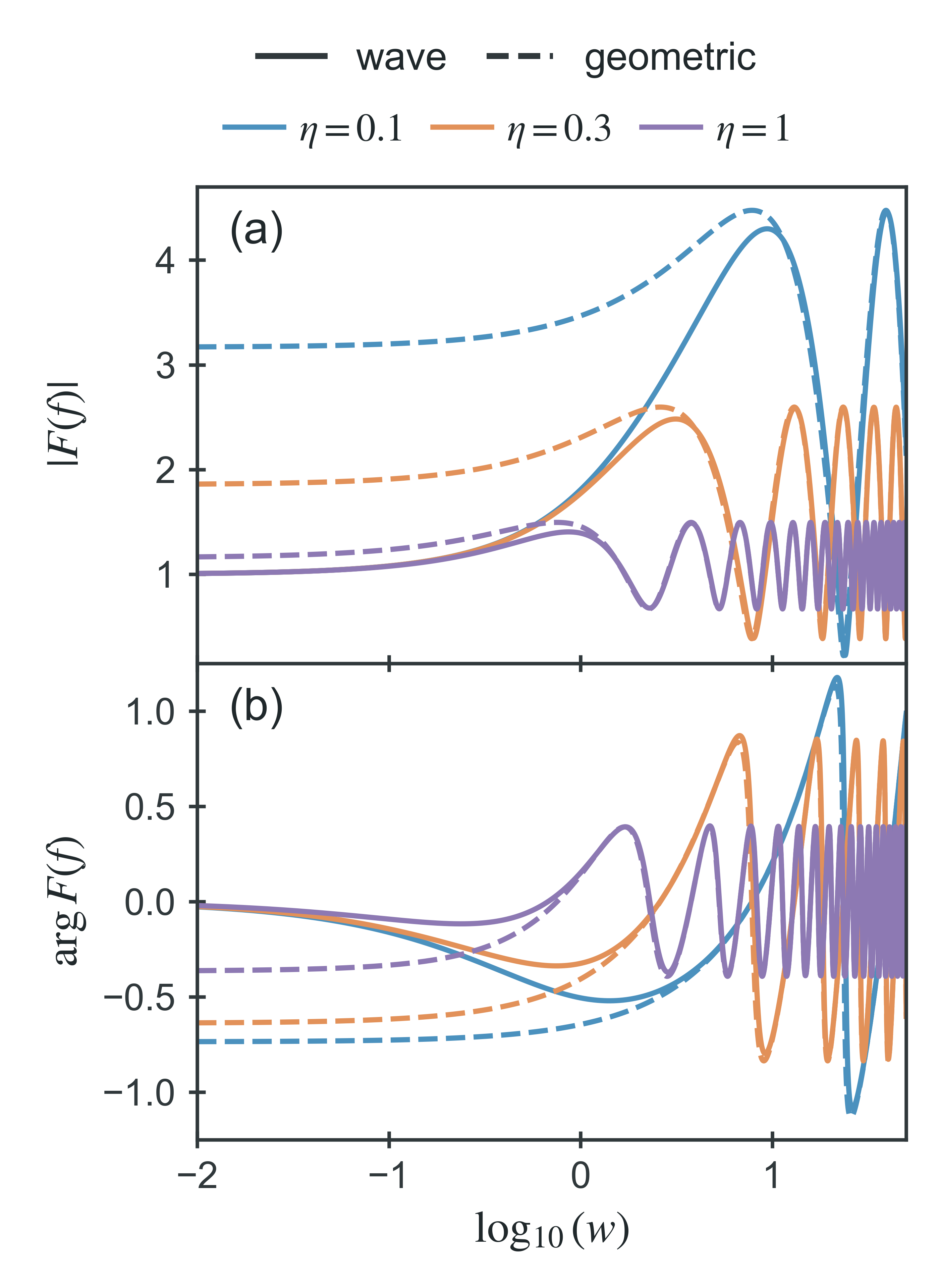}
\caption{Point-mass wave-optics and geometric-optics amplification for $\eta=0.1$ (blue), $0.3$ (orange), and $1$ (purple). Solid curves show wave-optics results, and dashed curves show geometric-optics results. In (a) and (b), we show the magnitude $|F|$ of the amplification factor and its principal-value phase $\arg F$, respectively.}
\label{fig:lens_amplification}
\end{figure}

For $w\gg1$ and sufficiently large phase separation between the images, the wave-optics expression approaches the coherent two-image geometric-optics limit~\cite{Takahashi2003Wave}. In the standard phase convention, the geometric-optics amplification takes the form
\begin{equation}
F_{\rm geo}=\sqrt{\mu_+}-i\sqrt{|\mu_-|}\exp(2\pi if\Delta t_l).
\label{eq:lens_geo}
\end{equation}
The signed image magnifications are
\begin{equation}
\mu_\pm=\frac{1}{2}\pm\frac{\eta^2+2}{2\eta\sqrt{\eta^2+4}},
\end{equation}
with $\mu_-<0$ for the secondary image. The observer-frame delay between the images is
\begin{equation}
\Delta t_l=4M_L\left[\frac{\eta\sqrt{\eta^2+4}}{2}+\ln\!\left(\frac{\sqrt{\eta^2+4}+\eta}{\sqrt{\eta^2+4}-\eta}\right)\right].
\end{equation}
This delay determines the relative image phase
\begin{equation}
\Delta\Phi_{\rm img}=2\pi f\Delta t_l,
\label{eq:lens_image_phase}
\end{equation}
which we use to distinguish the wave- and geometric-optics regimes~\cite{NakamuraDeguchi1999Wave}. We adopt $\Delta\Phi_{\rm img}=6$ as the switching value, based on numerical comparisons of the two expressions over the parameter range considered in this work. Within each lensing event, the amplification is $F_{\rm wave}$ for frequency components with $\Delta\Phi_{\rm img}<6$ and $F_{\rm geo}$ otherwise. Different components of the same event can therefore fall into different regimes.

As shown in Fig.~\ref{fig:lens_amplification}, the wave- and geometric-optics results differ at small $w$ and approach each other as the relative image phase increases. The transition depends on source position through the image time delay, motivating $\Delta\Phi_{\rm img}$ rather than $w$ alone as the switching diagnostic. The geometry and amplification defined above apply at a given source position and GW frequency. We apply these expressions event by event to the detector-frame signal.

\subsection{Event-local waveform construction}
\label{subsec:event_local_waveform}

\begin{figure*}[!t]
\centering
\includegraphics[width=0.96\textwidth]{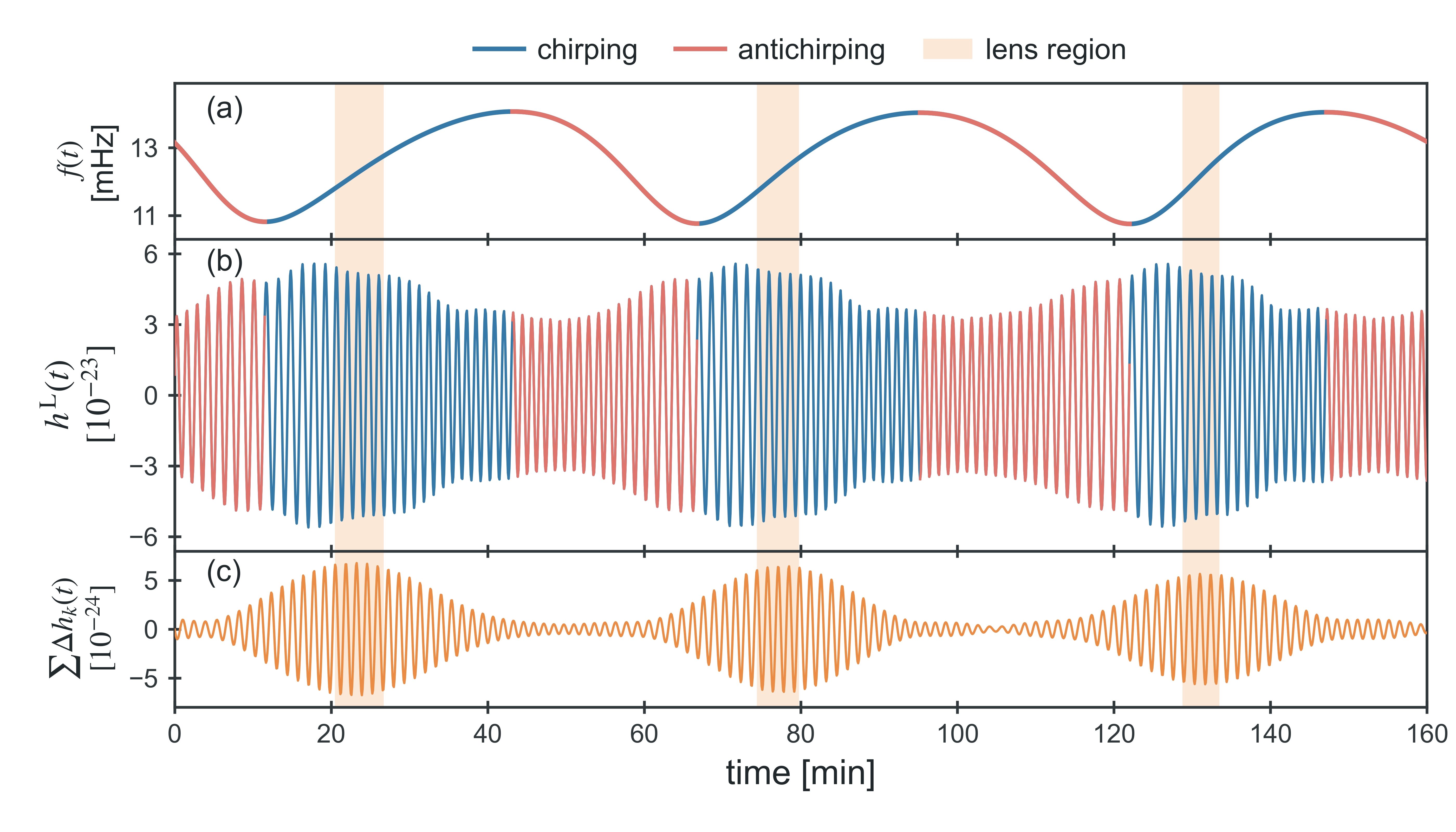}
\caption{Illustrative waveform with moving-source modulation and repeated lensing. (a) shows the detector-frame GW frequency $f(t)$, with chirping and antichirping intervals marked in blue and red. (b) shows the lensed waveform $h^{\rm L}(t)$, and (c) shows the total lensing correction $\sum_k\Delta h_k(t)$. The shaded regions indicate the lensing regions. The source parameters are $f_0\simeq12.96\,\mathrm{mHz}$, $m_1=m_2=110\,\mathrm{M}_\odot$, $m_3=2\times10^5\,\mathrm{M}_\odot$, $a_{\rm o}=32R_s$, $e_{\rm o}=0.3$, $\chi_3=0$, and $D_L\simeq51.5\,\mathrm{Mpc}$.}
\label{fig:event_local_frequency}
\end{figure*}

The intrinsic inspiral frequency of the inner BBH generally increases, but outer-orbit motion and the moving-source time transformation can produce intervals of decreasing detector-frame frequency $f(t)$~\cite{Toubiana2021Environmental}. A hierarchical triple can satisfy the lensing condition several times during an observation. The resulting $f(t)$ can be nonmonotonic over the full signal, so a detector-frame frequency may occur at more than one time. A global stationary-phase construction based on a unique inverse $t(f)$ is therefore not directly applicable. Each lensing event is instead treated using its local frequency evolution (see Fig.~\ref{fig:event_local_frequency}).

For the $k$th lensing event, let $\mathcal I_k=[t_{k,s},t_{k,e}]$ denote its core interval and define $f_k(t)=f(t)|_{\mathcal I_k}$. The interval boundaries are located continuously from the lensing geometry. For monotonic frequency evolution within this interval, we use the event-local inverse
\begin{equation}
t_k(f)=f_k^{-1}(f).
\label{eq:event_local_inverse}
\end{equation}
We construct this inverse by numerical interpolation over the event frequency range. The resulting amplification is
\begin{equation}
F_k(f)=F\!\left[f,t_k(f)\right].
\end{equation}
At the corresponding event time, we evaluate $F$ using Eqs.~\eqref{eq:lens_wave} and \eqref{eq:lens_geo} according to the switching prescription described above. The local inverse $t_k(f)$ supplies the instantaneous lensing geometry and relative image phase for each frequency component.

To isolate the waveform near each event without a sharp temporal cut, we use the Planck smooth step~\cite{McKechan2010Taper}
\begin{equation}
S(x)=\begin{cases}
0, & x\leq 0,\\
\displaystyle\frac{1}{1+\exp\!\left(\frac{1}{x}-\frac{1}{1-x}\right)}, & 0<x<1,\\
1, & x\geq 1.
\end{cases}
\end{equation}
The smooth endpoints reduce the spectral leakage caused by a sharp cut. With $\tau_k=0.3(t_{k,e}-t_{k,s})$ as the transition width on each side, the event window becomes
\begin{equation}
W_k(t)=S\!\left[\frac{t-(t_{k,s}-\tau_k)}{\tau_k}\right]S\!\left[\frac{(t_{k,e}+\tau_k)-t}{\tau_k}\right].
\end{equation}
The window takes the value 1 throughout the core interval and smoothly falls to 0 on either side. We apply it to the full unlensed detector-frame waveform $h(t)$ to obtain the local signal $h_k(t)=W_k(t)h(t)$.

We denote the Fourier transform of $h_k(t)$ by $\tilde h_k(f)$. Lensing changes this local spectrum by
\begin{equation}
\Delta\tilde h_k(f)=[F_k(f)-1]\tilde h_k(f).
\end{equation}
For the time-domain construction, we define the complementary weight $W_0(t)\equiv1-\sum_kW_k(t)$. It follows that $W_0(t)+\sum_kW_k(t)=1$ and $h(t)=W_0(t)h(t)+\sum_kW_k(t)h(t)$. Replacing each event-local contribution by its lensed counterpart gives
\begin{equation}
\begin{aligned}
h^{\rm L}(t)={}&W_0(t)h(t)
+\sum_k\mathcal F^{-1}\!\left\{F_k(f)\mathcal F[W_k(t)h(t)]\right\}\\
={}&h(t)+\sum_k\mathcal F^{-1}\!\left\{[F_k(f)-1]\mathcal F[W_k(t)h(t)]\right\}.
\end{aligned}
\end{equation}
The inverse-transform term in the second line is the event-local correction $\Delta h_k(t)$. Adding the correction from each event to the unlensed waveform $h(t)$ gives the lensed waveform
\begin{equation}
h^{\rm L}(t)=h(t)+\sum_k\Delta h_k(t).
\label{eq:event_local_lensed_waveform}
\end{equation}
The factor $F_k-1$ excludes the unlensed contribution already present in $h(t)$ within each event interval. When no lensing events occur, the event sum vanishes and $h^{\rm L}(t)=h(t)$. Each event correction is retained across the observation interval, including its tails outside the event window.

Figure~\ref{fig:event_local_frequency} illustrates the frequency evolution and waveform construction of a system undergoing repeated lensing. The Doppler effect from the outer-orbit motion modulates the detector-frame GW frequency, producing alternating chirping and antichirping intervals. The moving source also modulates the waveform amplitude through aberration, which changes the detector response. As the source moves along its outer orbit, it repeatedly enters the lensing region, giving rise to a sequence of lensing events. Within each event, the local inverse in Eq.~\eqref{eq:event_local_inverse} associates each frequency component with its corresponding time and lensing geometry. The resulting correction is transformed back to the time domain and added to the unlensed waveform, giving the full lensed waveform in Eq.~\eqref{eq:event_local_lensed_waveform}. In Fig.~\ref{fig:event_local_frequency}(c), the correction is strongest near the lensing events and becomes smaller between them. After the inverse transform, each event correction can extend beyond the shaded interval and is retained throughout the observation.

\section{Repeated lensing in evolving outer orbits}
\label{sec:repeated_lensing}

\subsection{Orbital evolution and repeated-lensing geometry}
\label{subsec:orbital_lensing_geometry}

\begin{figure*}[t]
\centering
\includegraphics[width=0.98\textwidth]{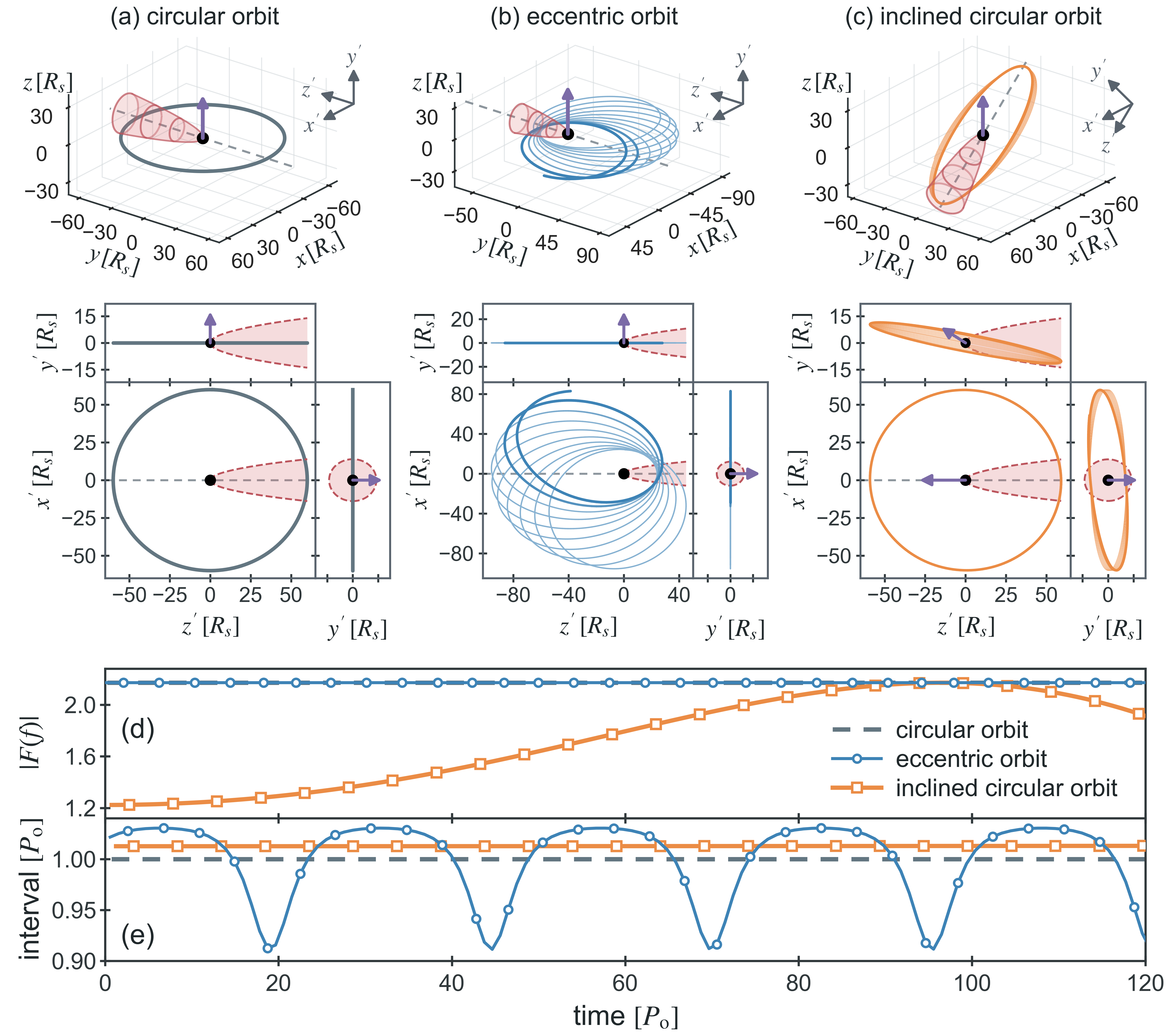}
\caption{Representative outer-orbit evolution and repeated lensing. Columns (a)--(c) show a circular equatorial orbit, an eccentric equatorial orbit, and an inclined circular orbit around a spinning SMBH, respectively. The upper part shows the trajectories over the first $10P_{\rm o}$ in three dimensions and projected along the three coordinate axes. $P_{\rm o}\simeq2\pi\sqrt{a_{\rm o}^3/m_3}$ is the reference Keplerian outer-orbit period. The $x'y'z'$ axes are indicated in the three-dimensional views. Black dots mark the SMBH, gray dashed lines the optical axis, and red shading the lensing region. Purple arrows indicate the SMBH spin axis. In (d) and (e), we show the peak amplification $|F|$ in each event and the interval between neighboring lensing-event peaks. The common parameters are $m_3=4\times10^6\,\mathrm{M}_\odot$, $a_{\rm o}=60R_s$, and a fixed frequency $f=3\,\mathrm{mHz}$. The settings are (a) $e_{\rm o}=\chi_3=0$, (b) $e_{\rm o}=0.6$ and $\chi_3=0$, and (c) $e_{\rm o}=0$, $\chi_3=0.99$, $i=70^\circ$, and $\epsilon=30^\circ$.}
\label{fig:orbit_evolution}
\end{figure*}

Repeated lensing depends on the geometry at conjunction and on how the outer orbit evolves over successive revolutions. Eccentric motion and relativistic periapsis precession alter the orbital phases and timing of conjunctions, while frame dragging around a spinning SMBH can reorient a non-equatorial orbital plane. The amplification and intervals between events can therefore vary from one orbit to the next. We examine these effects through the representative examples in Fig.~\ref{fig:orbit_evolution}.

Figure~\ref{fig:orbit_evolution}(a)--(c) shows the positions of the circular equatorial, eccentric equatorial, and inclined circular orbits relative to the SMBH, optical axis, and lensing region during the first $10P_{\rm o}$. Both equatorial orbits retain a fixed plane during the simulated evolution. The circular geometry repeats regularly, whereas the eccentric orbit undergoes periapsis precession~\cite{Laeuger2024Eccentric}. For the non-equatorial orbit around a Kerr SMBH, frame dragging induces LT precession and changes the orbital-plane orientation relative to the optical axis~\cite{Li2026LT}. This reorientation can move the orbit into or out of the lensing region as its alignment with the optical axis evolves. Such changes in lensing visibility distinguish non-equatorial motion around a spinning SMBH from motion in a fixed equatorial plane.

The event-by-event peak amplification in Fig.~\mbox{\ref{fig:orbit_evolution}}(d) remains nearly constant and is the same for the two equatorial cases. At the fixed GW frequency, both orbits repeatedly reach nearly exact alignment with the optical axis. For the inclined orbit, the amplification rises and falls as LT precession changes its projected alignment with the optical axis. An edge-on geometry generally favors a smaller projected source--lens separation and hence stronger lensing. In this example, the inclined orbit usually produces weaker amplification, but its peaks approach those of the equatorial orbits as the projected alignment improves.

The time intervals between the peaks of neighboring lensing events differ among the three orbits in Fig.~\ref{fig:orbit_evolution}(e). The circular equatorial reference has regular intervals, while periapsis precession changes the orbital phases of successive conjunctions in the eccentric case and produces periodic interval modulation. The Schwarzschild periapsis-precession period is~\cite{Naoz2016EKL}
\begin{equation}
T_{\rm prec}=\frac{2\pi a_{\rm o}^{5/2}(1-e_{\rm o}^2)}{3m_3^{3/2}}.
\end{equation}
For the eccentric configuration in Fig.~\ref{fig:orbit_evolution}(b), this gives $T_{\rm prec}\simeq25.6P_{\rm o}$, consistent with the modulation timescale visible in Fig.~\ref{fig:orbit_evolution}(e). The inclined circular example has nearly constant event intervals that are slightly longer than those of the circular equatorial reference. In these representative configurations, periapsis precession modulates the event intervals, whereas LT orbital-plane precession changes the peak amplification. Both effects can coexist on a general non-equatorial eccentric orbit and jointly change the amplitude and timing of repeated lensing.

\begin{figure}[!ht]
\centering
\includegraphics[width=0.95\columnwidth]{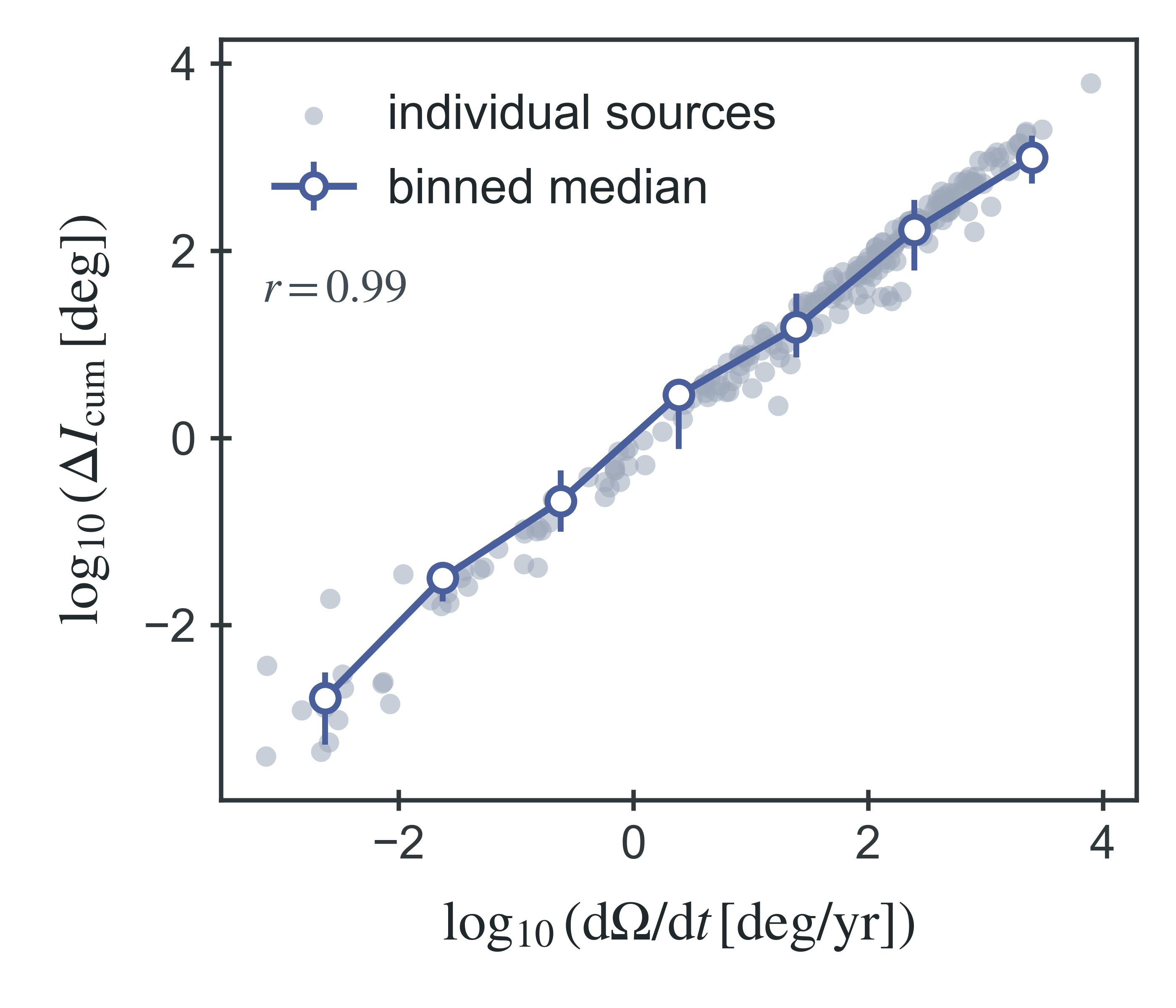}
\caption{Cumulative line-of-sight angular change for selected lensed systems with $\mathrm{SNR}\geq8$ over a one-year observation. The horizontal axis shows the LT nodal-precession rate in Eq.~\eqref{eq:lt_characteristic_rate}, while the vertical axis shows $\Delta I_{\rm cum}$. The latter is the cumulative angular change of $I$ over the observation, obtained by summing the absolute changes between successive samples. The points denote individual systems. Blue markers show the binned medians, and error bars span the 16th--84th percentiles.}
\label{fig:inclination_sweep}
\end{figure}

\begin{figure}[!ht]
\centering
\includegraphics[width=0.95\columnwidth]{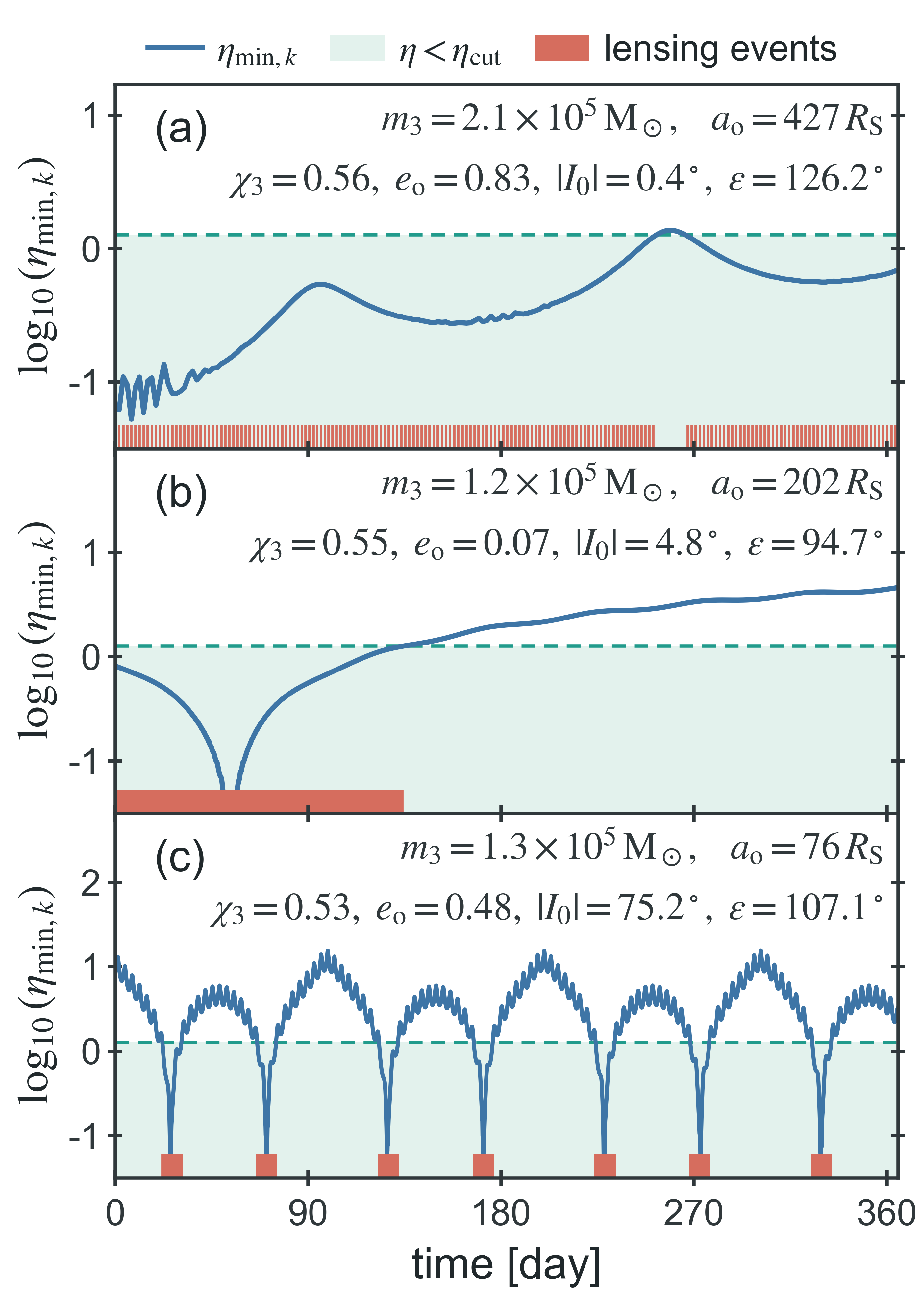}
\caption{Evolution of lensing geometry and visibility over one year. Panels (a), (b), and (c) show three representative lensed systems, with their parameters indicated in each panel. Blue curves connect the sampled minimum source positions $\eta_{\min,k}$ at successive conjunctions $k$. Dashed lines mark $\eta_{\rm cut}$, and green shading indicates $\eta<\eta_{\rm cut}$. Red marks indicate the occurrence of individual lensing events. The width of each red mark is chosen for visibility and does not represent the event duration.}
\label{fig:lensing_visibility}
\end{figure}

To quantify the orbital-plane reorientation induced by LT precession, we use the signed angle between the line of sight and the instantaneous outer-orbital plane,
\begin{equation}
\begin{gathered}
\sin I(t)=\hat{\mathbf L}_{\rm o}(t)\cdot\hat{\mathbf N}\\
=\cos i(t)\cos\epsilon-\sin i(t)\sin\epsilon\cos\Omega(t).
\end{gathered}
\end{equation}
The edge-on configuration corresponds to $I=0$, and LT precession makes $I(t)$ evolve from its initial value $I_0=I(0)$. We compare the numerically accumulated change of $I$ with the weak-field LT nodal-precession rate~\cite{Li2026LT},
\begin{equation}
\frac{\mathrm d\Omega}{\mathrm dt}=\frac{2\chi_3m_3^2}{a_{\rm o}^3(1-e_{\rm o}^2)^{3/2}}.
\label{eq:lt_characteristic_rate}
\end{equation}
This expression gives the secular LT rate of the node angle $\Omega$ from the source parameters. The numerical evolution follows the full adopted PN dynamics, so changes in $I$ need not match this secular estimate exactly.

The weak-field LT rate and the numerical cumulative change of $I$ in Fig.~\ref{fig:inclination_sweep} have a log-space Pearson correlation of $r=0.99$ in the selected sample. The binned medians increase with the LT rate, indicating that precession sets the characteristic scale for accumulated orbital-plane reorientation. The magnitude of this change also depends on orbital configuration and viewing geometry.

The lensing sequences in \mbox{Fig.~\ref{fig:lensing_visibility}} show how LT precession affects repeated lensing. As the outer orbital plane reorients, the minimum $\eta$ reached at successive conjunctions changes, opening or closing intervals of favorable alignment. These intervals show how long favorable alignment persists, whereas the event count also depends on the outer-orbit period.

The duration and timing of favorable alignment differ among the three systems. In \mbox{Fig.~\ref{fig:lensing_visibility}(a)}, alignment persists through most of the observation, with a brief interruption when the minimum $\eta$ exceeds the threshold. In contrast, the minimum $\eta$ first decreases and then increases in \mbox{Fig.~\ref{fig:lensing_visibility}(b)}, concentrating lensing in the early part of the observation. The system shown in \mbox{Fig.~\ref{fig:lensing_visibility}(c)} has an initial angle of $|I_0|\simeq75^\circ$. Orbital-plane precession repeatedly brings it into the lensing region, producing separated clusters of events.

\mbox{Figures~\ref{fig:orbit_evolution}--\ref{fig:lensing_visibility}} show that outer-orbit evolution modulates the occurrence, spacing, and amplification of repeated lensing events. Precession can bring initially unfavorable systems into the lensing region and move favorably aligned systems out of it. Assessing repeated lensing and its contribution to parameter inference therefore requires following the geometry throughout the observation.

\subsection{Lensing probability and event statistics}
\label{subsec:lensing_probability_statistics}

The circular-orbit lensing probability in Ref.~\cite{DOr2020Repeated} provides a simple reference for estimating how the outer-orbit geometry affects repeated lensing. A circular orbit has a fixed radius at superior conjunction, when the BBH lies behind the SMBH relative to the observer. For an eccentric orbit, the separation at this alignment varies with orbital phase, changing the local Einstein scale and the allowed viewing-angle range. We extend the circular expression to obtain an approximate geometric lensing probability, with the derivation and orientation-prior dependence given in Appendix~\ref{app:eccentric_lensing_probability}. This estimate describes geometric trends rather than the fraction of sources in the selected numerical sample.

For a reference Keplerian ellipse, the source--SMBH separation at this alignment is
\begin{equation}
r_c=\frac{a_{\rm o}(1-e_{\rm o}^2)}{1+e_{\rm o}\cos[\nu_c]},
\end{equation}
where $\nu_c$ is the corresponding true anomaly measured from periapsis. The leading geometric correction relative to the circular orbit is
\begin{equation}
\mathcal E(e_{\rm o})
=\left\langle\sqrt{\frac{a_{\rm o}}{r_c}}\right\rangle_{\nu_c}
=\frac{2E(k)}{\pi\sqrt{1-e_{\rm o}}},
\label{eq:eccentric_geometric_factor}
\end{equation}
with $k^2=2e_{\rm o}/(1+e_{\rm o})$. The brackets $\langle\cdots\rangle_{\nu_c}$ denote a uniform average over $\nu_c\in[0,2\pi)$, and $E(k)$ is the complete elliptic integral of the second kind. The limit $\mathcal E(0)=1$ recovers the circular result. The elliptic-integral expression is exact for this uniform-$\nu_c$ average. Its substitution into the nonlinear angular probability remains approximate.

The resulting lensing-probability estimate is
\begin{equation}
P_{\rm L}\simeq\min\!\left[\frac{T_{\rm obs}}{2P_{\rm o}},1\right]\frac{2}{\pi}\arcsin\!\left[\min\!\left[1,2\eta_{\rm cut}\sqrt{\frac{m_3}{a_{\rm o}}}\mathcal E(e_{\rm o})\right]\right].
\label{eq:eccentric_lensing_opportunity}
\end{equation}
The observation time is denoted by $T_{\rm obs}$, and $\min[a,b]$ denotes the smaller of $a$ and $b$. The inner minimum keeps the argument of $\arcsin$ within its allowed range, while the outer minimum represents the heuristic finite-observation prescription adopted from Ref.~\cite{DOr2020Repeated}.

\begin{figure}[!ht]
\centering
\includegraphics[width=0.96\columnwidth]{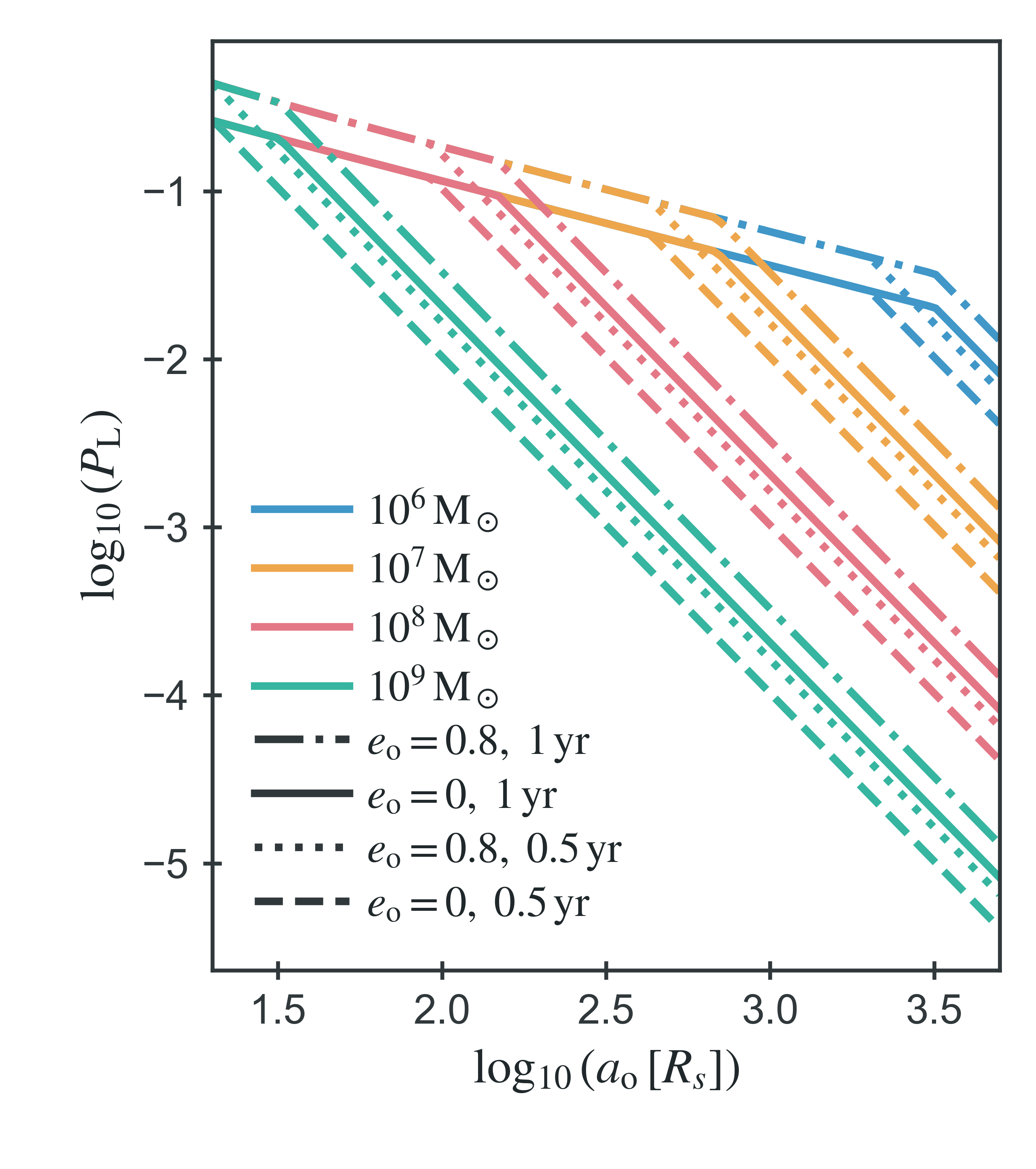}
\caption{Approximate lensing probability $P_{\rm L}$ in Eq.~\eqref{eq:eccentric_lensing_opportunity}. Colors distinguish $m_3=10^6$, $10^7$, $10^8$, and $10^9\,\mathrm{M}_\odot$. Line styles distinguish $e_{\rm o}=0$ and $0.8$ for $T_{\rm obs}=0.5$ and $1\,{\rm yr}$, as indicated in the legend.}
\label{fig:lensing_probability}
\end{figure}

\begin{figure}[!ht]
\centering
\includegraphics[width=0.96\columnwidth]{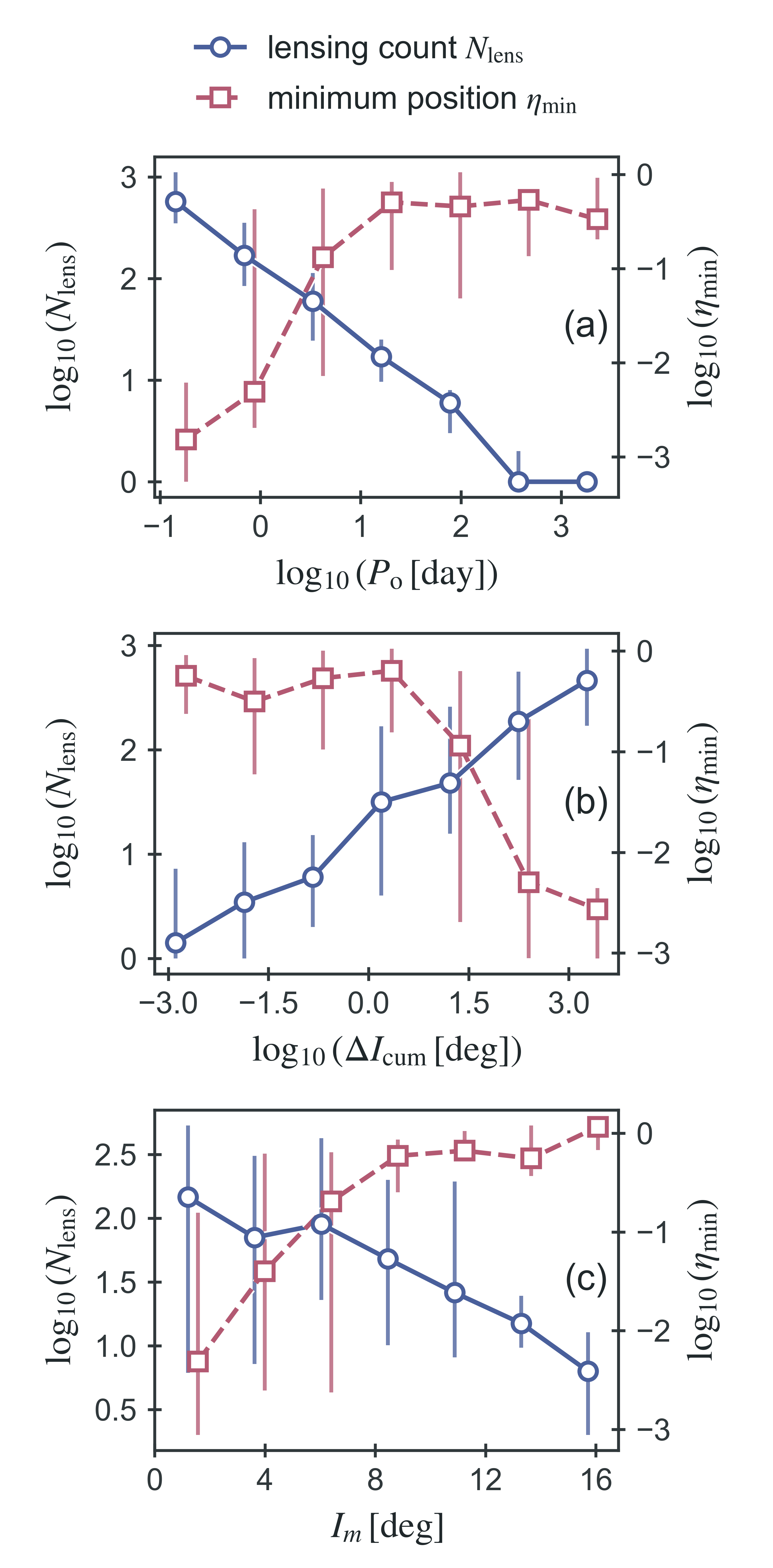}
\caption{Repeated-lensing event statistics for selected lensed systems over one year. (a) shows the event statistics as functions of $P_{\rm o}$. (b) shows their dependence on $\Delta I_{\rm cum}$. (c) shows their dependence on $I_m$. The angle $I_m$ is the absolute value of the median $I$ evaluated at the lensing-event peaks of each system. On the left and right axes, $N_{\rm lens}$ is the number of identified lensing events during the one-year observation, and $\eta_{\min}$ is the minimum source position among all lensing events of a system. Markers and error bars follow the statistical convention in Fig.~\ref{fig:inclination_sweep}.}
\label{fig:lensing_event_statistics}
\end{figure}

The approximate probability in Fig.~\ref{fig:lensing_probability} decreases with increasing $a_{\rm o}$ because a larger separation narrows the allowed viewing-angle range and lengthens $P_{\rm o}$, reducing the orbital passages sampled within a fixed observation. Increasing $e_{\rm o}$ raises the probability in the adopted eccentric approximation through the growth of $\mathcal E(e_{\rm o})$, which enlarges the phase-averaged angular lensing range. A longer $T_{\rm obs}$ samples more of the outer motion and increases the chance of favorable alignment until the finite-observation factor saturates. At the same separation in units of $R_s$, a larger $m_3$ implies a larger physical orbit and a longer $P_{\rm o}$, strengthening the finite-observation suppression.

The approximate probability above captures the averaged geometric dependence but does not follow individual lensing events along a PN-evolved orbit. We therefore use the selected lensed systems defined in \mbox{Sec.~\ref{sec:triple_signal}} to examine how the orbital timescale and time-dependent viewing geometry shape the repeated-event statistics.

\mbox{Figure~\ref{fig:lensing_event_statistics}(a)} shows that the event count $N_{\rm lens}$ decreases with increasing outer-orbit period $P_{\rm o}$, while $\eta_{\min}$ generally increases. Shorter periods provide more conjunctions within one year and more opportunities for close alignment, consistent with the larger event counts and smaller $\eta_{\min}$. In logarithmic space, the event count has a Pearson correlation of $r=-0.94$ with $P_{\rm o}$ in the selected sample.

\mbox{Figure~\ref{fig:lensing_event_statistics}(b)} shows that $N_{\rm lens}$ increases with the cumulative angular change $\Delta I_{\rm cum}$, with a log-space Pearson correlation of $r=0.76$. Meanwhile, $\eta_{\min}$ decreases markedly toward the large-$\Delta I_{\rm cum}$ end. A larger cumulative angular change reflects greater variation in the viewing geometry, which can alter the alignment at successive conjunctions. Such evolution can provide more lensing opportunities and bring some conjunctions into closer alignment, consistent with the larger event counts and smaller $\eta_{\min}$.

\mbox{Figure~\ref{fig:lensing_event_statistics}(c)} relates the orbital orientation at lensing-event peaks to both statistics. A small $I_m$ can arise when event-peak inclinations are close to zero or when positive and negative inclinations approximately balance. The binned medians generally show fewer lensing events and larger $\eta_{\min}$ as $I_m$ increases.

These trends reflect the coupled effects of orbital timescale and evolving viewing geometry. The orbital period sets the recurrence timescale of conjunctions, while orbital-plane evolution determines whether they reach favorable alignment. We next test whether a circular, nonprecessing outer orbit identifies lensing in the systems lensed by the full evolving model.
\subsection{Comparison with a circular outer-orbit baseline}
\label{subsec:circular_baseline}

\begin{figure}[!ht]
\centering
\includegraphics[width=0.96\columnwidth]{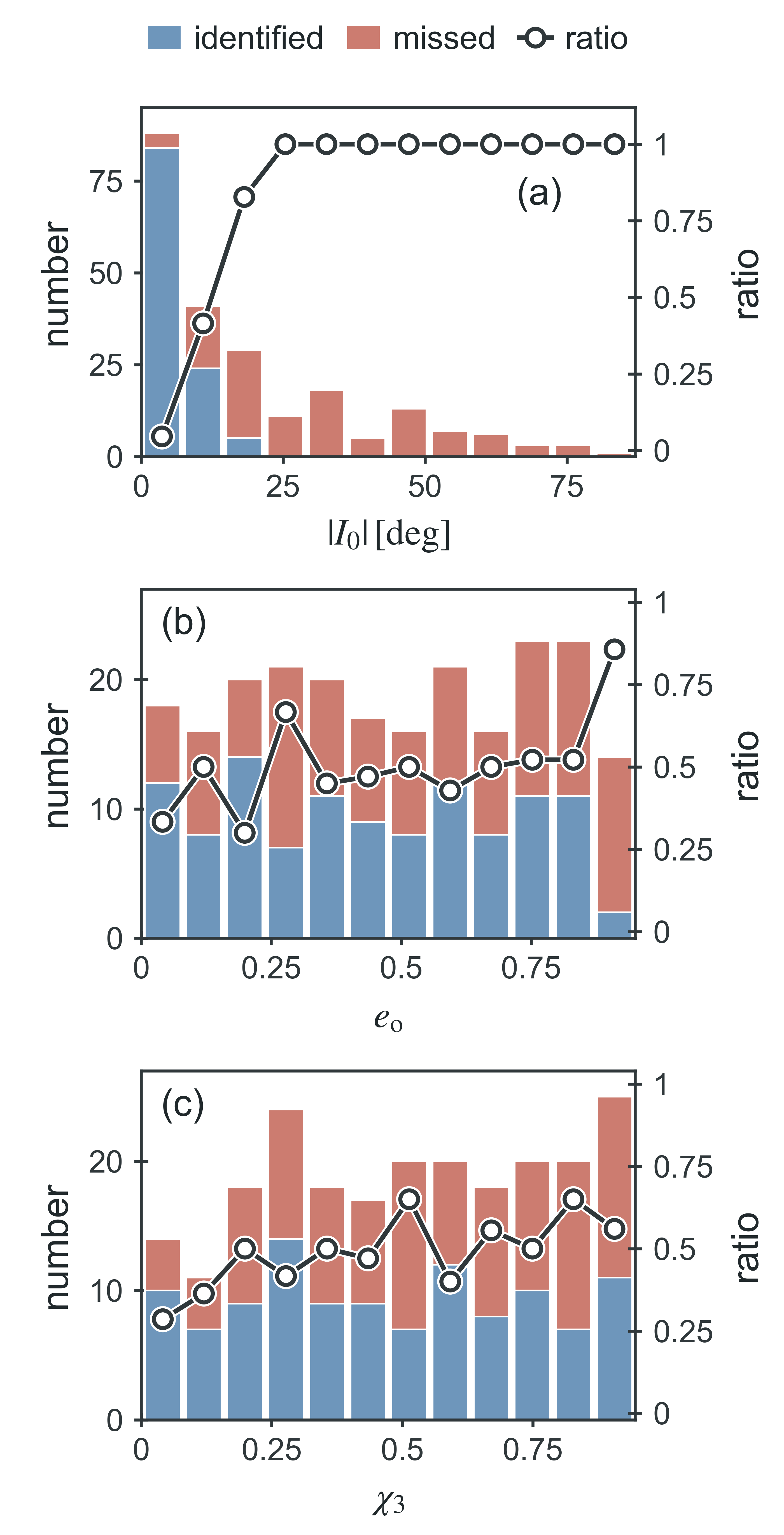}
\caption{Lensing identification with a circular, nonprecessing outer-orbit baseline for systems lensed in the full evolving model. Panels (a), (b), and (c) show the dependence on the initial angle $|I_0|$, eccentricity $e_{\rm o}$, and SMBH spin $\chi_3$. Blue bars count systems retaining at least one lensing event in the baseline, while red bars count those with no event. Dark gray curves show the fraction of systems missed in each bin.}
\label{fig:circular_baseline_missed}
\end{figure}

For the selected systems with lensing in the full evolving model, we use a simplified Keplerian circular and nonprecessing outer orbit as a baseline. The inner BBH waveform, masses, outer semimajor-axis input, initial geometry, detector response, and one-year observation time remain unchanged. Using the same lensing criterion, we classify systems as missed if the circular baseline produces no event during the one-year observation, and as retained otherwise. The fixed circular geometry repeats each orbit, so one period suffices to screen for the absence of events. The comparison in \mbox{Fig.~\ref{fig:circular_baseline_missed}} uses the selected lensed systems defined above.

The circular baseline misses all lensing events for 112 of the 225 selected systems, corresponding to $49.8\%$. The missed fraction rises substantially with $|I_0|$ in Fig.~\ref{fig:circular_baseline_missed}(a). Systems initially close to edge-on are more often retained because even a fixed Keplerian circular plane can readily satisfy the lensing geometry. At larger $|I_0|$, LT-driven orbital-plane precession in the full model can change the viewing geometry during the observation, bringing an initially unfavorable orbit closer to edge-on and allowing lensing events to occur. Without this evolution, the fixed circular baseline may miss these systems. In the current sample, all systems in bins above $|I_0|\simeq22^\circ$ are missed by the circular baseline.

The missed fraction shows a modest overall increase with both $e_{\rm o}$ and $\chi_3$ in \mbox{Fig.~\ref{fig:circular_baseline_missed}(b)} and \mbox{(c)}. Both trends exhibit substantial bin-to-bin fluctuations and are weaker than the dependence on the initial angle. Higher eccentricity changes the source--lens separation at conjunction and the temporal distribution of successive alignments. Larger SMBH spin enhances LT precession, allowing more extensive orbital-plane evolution during the observation. These changes can bring systems into lensing configurations that the fixed circular baseline does not reach, increasing the missed fraction. Both effects act together with the initial viewing geometry, so the comparison reflects the combined differences between the full evolving model and the circular, nonprecessing baseline.

The circular baseline can thus miss a lensed source entirely, in addition to changing the predicted event times and amplification. Following outer-orbit evolution helps identify systems that enter the lensing region during the observation and preserves their associated waveform features. Unlike this orbital-baseline comparison, the analysis in Sec.~\ref{sec:inference} retains the full evolving outer-orbit model and switches lensing on or off to examine its effect on parameter inference.
\section{Parameter constraints and waveform residuals}
\label{sec:inference}

\subsection{Impact of repeated lensing on parameter constraints}
\label{subsec:fisher_constraints}

We use a local Fisher-matrix analysis to compare parameter constraints with and without repeated lensing~\cite{CutlerFlanagan1994}. The Fisher information matrix is
\begin{equation}
\Gamma_{ij}=\left(\frac{\partial h}{\partial\theta_i}\middle|\frac{\partial h}{\partial\theta_j}\right).
\end{equation}
We compute the Fisher matrix from numerical waveform derivatives, using finite-difference steps selected through stability tests. The matrix is accumulated in a numerically stable basis and inverted using extended precision. Its inverse gives the covariance $\Sigma=\Gamma^{-1}$, with one-sigma uncertainties $\sigma_{\theta_i}=\sqrt{\Sigma_{ii}}$. Both matrices are transformed back to the original parameter coordinates. Inversion consistency is checked through $\epsilon_{\rm inv}=\max_{ij}|(\Gamma\Sigma-\mathbb I)_{ij}|$, and only calculations with $\epsilon_{\rm inv}\le10^{-3}$ are retained. This residual is evaluated using the paired matrices at their retained precision.

\begin{figure}[!ht]
\centering
\includegraphics[width=0.98\columnwidth]{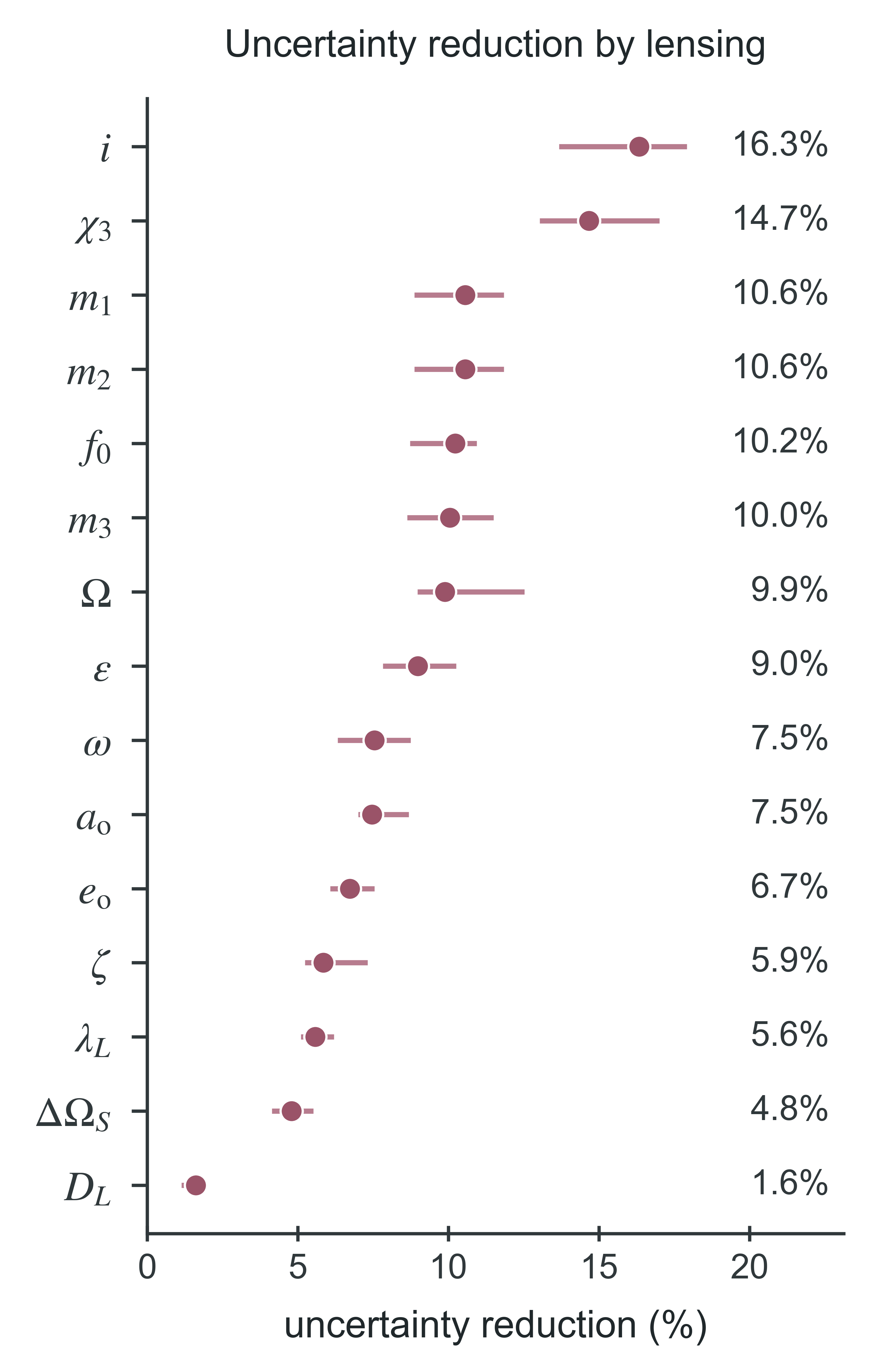}
\caption{Reduction in parameter uncertainties due to lensing for selected lensed systems. For each system, the reduction is evaluated relative to its unlensed uncertainty. Circles show the medians of these individual-source percentage reductions. Horizontal error bars indicate 68\% confidence intervals for the medians from paired bootstrap resampling of systems. Percentages on the right give the corresponding medians.}
\label{fig:fisher_uncertainty_comparison}
\end{figure}

The Fisher-matrix calculation uses 16 parameters,
\begin{equation}
\begin{aligned}
\boldsymbol{\theta}=\{&
\ln f_0,\ln m_1,\ln m_2,\ln m_3,\ln D_L,\\
&\ln a_{\rm o},e_{\rm o},i,\Omega,\omega,\chi_3,\Theta,\Phi,\epsilon,\zeta,\lambda_L\}.
\end{aligned}
\label{eq:fisher_parameters}
\end{equation}
The lensed and unlensed calculations share the same source parameters, observation time, LISA response, noise model, and parameter space. All waveform ingredients remain identical except for the lensing correction. The sky angles $\Theta$ and $\Phi$ are included in the Fisher matrix, and we derive the sky-localization area from their covariance~\cite{Cutler1998LISA},
\begin{equation}
\Delta\Omega_S=2\pi|\sin\Theta|
\sqrt{\Sigma_{\Theta\Theta}\Sigma_{\Phi\Phi}-\Sigma_{\Theta\Phi}^2}.
\end{equation}
We compare the lensed and unlensed parameter uncertainties for the selected lensed systems in Fig.~\ref{fig:fisher_uncertainty_comparison}.

\mbox{Figure~\ref{fig:fisher_uncertainty_comparison}} shows positive median percentage reductions in uncertainty for all displayed quantities, calculated from lensed and unlensed results for each source. The largest median reductions occur for the outer-orbit inclination $i$ and SMBH spin $\chi_3$, at approximately $16.3\%$ and $14.7\%$. Orbital orientation and spin jointly affect the evolving source--lens alignment, allowing successive lensing events to carry additional geometric information. The median reductions for the other outer-orbit and orientation parameters $\Omega$, $\epsilon$, $a_{\rm o}$, and $e_{\rm o}$ are approximately $9.9\%$, $9.0\%$, $7.5\%$, and $6.7\%$, respectively.

The median uncertainty reduction for the SMBH mass $m_3$ is approximately $10.0\%$. The lens mass sets the characteristic time delays and the frequency dependence of the lensing phase. The outer semimajor axis also affects the source--lens separation and the timing of repeated alignments. Lensing adds mass and geometric information to that provided by the orbital modulation in the unlensed signal. The component masses $m_1$ and $m_2$ each show a median reduction of approximately $10.6\%$. The median uncertainty reduction for $f_0$ is $10.2\%$, comparable to those for the mass parameters. By comparison, the median reductions in sky-localization area $\Delta\Omega_S$ and uncertainty in luminosity distance $D_L$ are smaller, at approximately $4.8\%$ and $1.6\%$.

Repeated lensing therefore supplies additional constraints on several mass, orbital, and orientation parameters. Orbital inclination and SMBH spin show the largest median improvements in the current individual-source comparison. We next examine whether the lensing-induced structure can be absorbed by parameter shifts when lensing is omitted from the analysis model.
\subsection{Residual signatures of neglected lensing}
\label{subsec:lensing_residual}

We compare the lensed reference signal $h^{\rm L}$ with the unlensed model $h^{\rm UL}$ evaluated at the same source parameters. Their difference is
\begin{equation}
\delta h=h^{\rm L}-h^{\rm UL}.
\end{equation}
The linearized waveform-model error formalism identifies the component of this difference that local variations of the unlensed-model parameters can absorb~\cite{CutlerVallisneri2007}. Using the local tangent-space projection~\cite{Vallisneri2012RealityCheck}, we evaluate the residual SNR as
\begin{equation}
\rho_{\rm res}^2=(\delta h|\delta h)-b_i\Sigma^{\rm UL}_{ij}b_j,
\end{equation}
with $b_i=(\partial_i h^{\rm UL}|\delta h)$ and $\Sigma^{\rm UL}=(\Gamma^{\rm UL})^{-1}$. The waveform derivatives and Fisher matrix are evaluated from the unlensed model at the reference source parameters. Repeated indices are summed over the 16 parameters in Eq.~\eqref{eq:fisher_parameters}.

\begin{figure}[!ht]
\centering
\includegraphics[width=0.98\columnwidth]{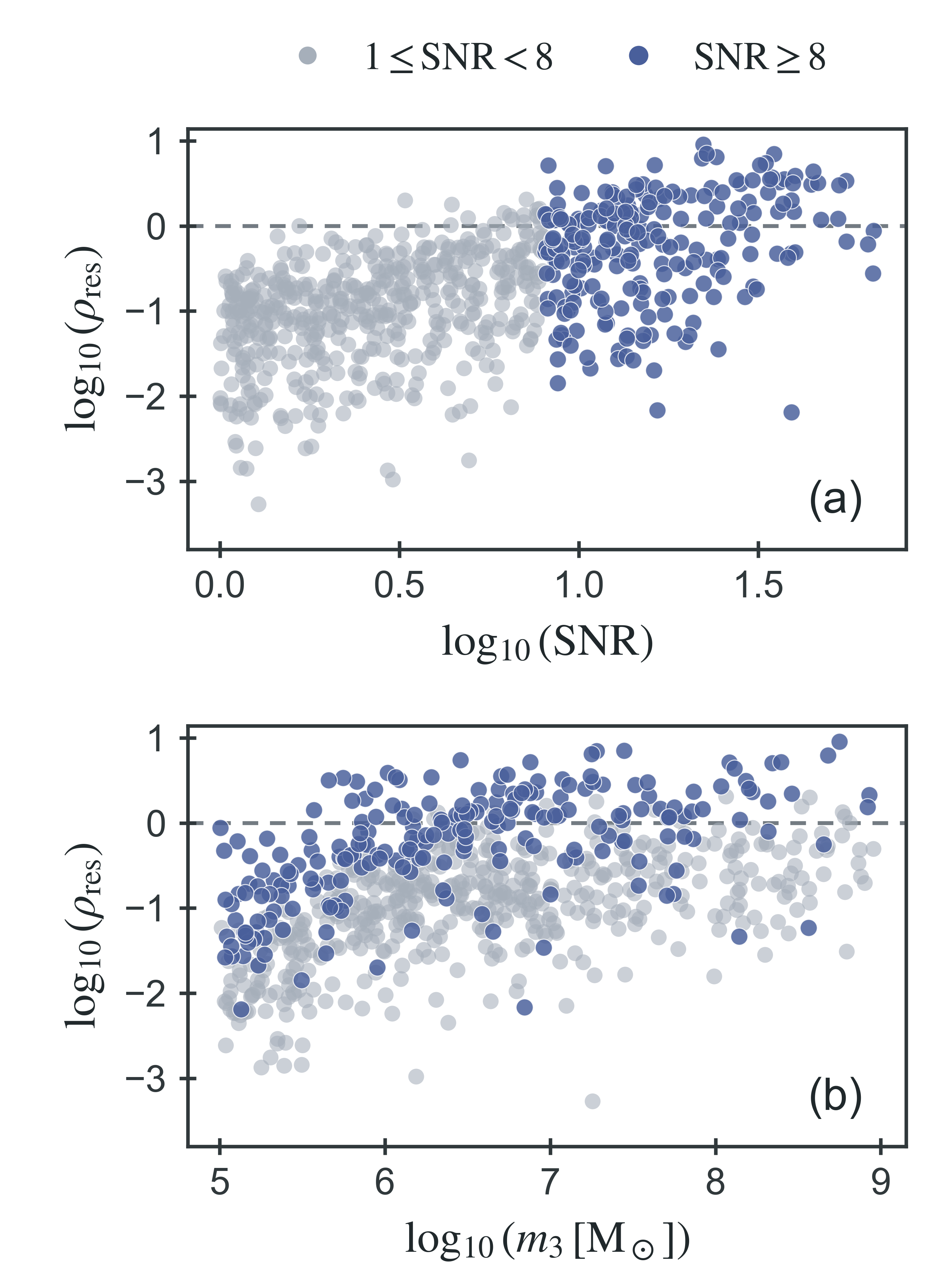}
\caption{Fisher-projected residual SNR for lensed systems. Panel (a) shows its dependence on source SNR, and panel (b) shows its dependence on SMBH mass $m_3$. Gray points denote systems with $1\leq\mathrm{SNR}<8$, while blue points denote systems with $\mathrm{SNR}\geq8$. Horizontal dashed lines mark $\rho_{\rm res}=1$.}
\label{fig:lensing_projected_residual}
\end{figure}

We evaluate the full 16-parameter projection using singular value decomposition and adopt \mbox{$\rho_{\rm res}=1$} as a reference level~\cite{Lindblom2008Accuracy}. \mbox{Figure~\ref{fig:lensing_projected_residual}} shows the projected residual for lensed systems with valid residual calculations and $\mathrm{SNR}\ge1$. With an adopted detection threshold of $\mathrm{SNR}=8$, $41.8\%$ of the systems meeting this threshold retain $\rho_{\rm res}\ge1$ after projection.

\mbox{Figure~\ref{fig:lensing_projected_residual}(a)} shows an overall increase in the projected residual with source SNR. For comparable fractional waveform differences, a higher-SNR signal makes the lensing-induced structure more prominent in the noise-weighted inner product. At fixed SNR, the scatter reflects differences in lensing geometry and in how much of the waveform difference local parameter shifts can absorb.

The projected residual tends to increase with SMBH mass in \mbox{Fig.~\ref{fig:lensing_projected_residual}(b)}, with substantial scatter. The lens mass sets the characteristic time-delay and frequency-dependent phase scales, shaping the difference between the lensed waveform and the unlensed model. The residual after local projection also depends on orbital configuration, viewing geometry, and signal strength.

For some systems meeting the adopted detection threshold, local adjustments of the unlensed-model parameters still leave appreciable lensing-induced waveform structure. Along with the uncertainty reductions found in Sec.~\ref{subsec:fisher_constraints}, this result shows that neglecting lensing can both discard useful parameter information and leave an unmodeled waveform component.
\section{Discussion and conclusions}
\label{sec:conclusion}

In this work, we investigate repeated gravitational lensing of GWs from a stellar-mass BBH on an eccentric, inclined, and relativistically evolving orbit around a spinning SMBH. Our model combines numerical PN outer-orbit dynamics with an inspiral waveform that includes source-motion and precession effects. An event-local lensing construction follows the changing source--lens geometry. For a one-year LISA observation, we examine lensing-event identification, parameter constraints with and without lensing, and residual waveform structure when lensing is neglected.

Our results show that outer-orbit evolution affects the identification of repeated lensing and the parameter information in the lensed waveform. Eccentric motion, relativistic periapsis precession, and orbital-plane evolution change successive source--lens alignments. The timing and amplification of successive events can vary, and systems can enter or leave favorable lensing configurations during the observation. A fixed circular and nonprecessing approximation misses $49.8\%$ of the selected lensed systems, with the missed fraction increasing markedly with $|I_0|$. This comparison tests the combined orbital model rather than isolating eccentricity or spin-induced precession.

Repeated lensing supplies additional mass and geometric information. The median individual-source uncertainty reductions are largest for the outer-orbit inclination $i$ and SMBH spin $\chi_3$, at $16.3\%$ and $14.7\%$, respectively. Lensing time delays and phase structure supplement the mass information from unlensed orbital modulation, while changes in source--lens separation and alignment provide geometric information. With an adopted detection threshold of $\mathrm{SNR}=8$, $41.8\%$ of the systems meeting this threshold retain $\rho_{\rm res}\ge1$ after the local Fisher projection. Including lensing in the waveform can recover additional parameter information and account for structure that remains after local adjustments of the unlensed model.

Our conclusions apply to standard repeated lensing under an event-local scalar point-mass model. Stronger-field and spin-dependent propagation lie beyond this model. The Fisher analysis estimates local parameter information, while the projected residual measures waveform structure remaining after local adjustments of the unlensed model. Full posterior inference and global nonlinear model comparison offer complementary tests beyond this local description. The reported fractions and trends refer to the selected simulated systems and the adopted one-year LISA configuration. They are conditional sample statistics rather than astrophysical population-rate predictions.

Bayesian parameter estimation with the present waveforms could test whether the posterior distributions reproduce the uncertainty trends predicted by the Fisher analysis. Nonlinear inference could test whether the residual persists after parameter reoptimization and assess the statistical distinction between lensed and unlensed models. Future waveform extensions could include inner-binary eccentricity and misaligned component spins with intrinsic spin precession. A more detailed treatment of the coupled inner and outer dynamics would test how these additional effects interact with repeated lensing. Future propagation models could include stronger-field lensing, SMBH-spin-dependent effects, retrolensing, and polarization dependence. These extensions would test how propagation beyond the scalar point-mass approximation affects lensing signatures and parameter constraints. Astrophysical population modeling would subsequently connect the conditional trends to the expected number of observable sources. Modeling orbital evolution and repeated lensing together can help constrain the SMBH and outer orbit in LISA observations of hierarchical BBHs.

\begin{acknowledgments}
This work was supported by the National Key Research and Development Program of China (Grant No. 2023YFC2206702), the National Natural Science Foundation of China (Grants No. 125B2102, No. 12575072, and No. 12547101), and the Fundamental Research Funds for the Central Universities Project (Grant No. 2024IAIS-ZD009).
\end{acknowledgments}

\appendix
\section{Lensing probability for an eccentric outer orbit}
\label{app:eccentric_lensing_probability}

This appendix derives the approximate lensing probability used in Sec.~\ref{subsec:lensing_probability_statistics} and specifies the orientation convention and phase-averaging prescription for eccentric orbits. For a reference Keplerian ellipse, the source--SMBH separation at alignment is
\begin{equation}
r_c=\frac{a_{\rm o}(1-e_{\rm o}^2)}{1+e_{\rm o}\cos[\nu_c]}.
\end{equation}
Near superior conjunction in a nearly edge-on orbit, the local line-of-sight separation and Einstein radius satisfy
\begin{equation}
D_{\rm LS,c}\simeq r_c,\qquad
R_{E,c}\simeq\sqrt{4m_3r_c}.
\end{equation}
The minimum transverse offset is approximately $u_{\min}\simeq r_c|\sin[I]|$. The condition $u_{\min}<\eta_{\rm cut}R_{E,c}$ gives
\begin{equation}
|\sin[I]|<2\eta_{\rm cut}\sqrt{\frac{m_3}{r_c}}.
\end{equation}
We write the allowed angular range using
\begin{equation}
\begin{aligned}
s(\nu_c)&=\min\!\left[1,2\eta_{\rm cut}\sqrt{\frac{m_3}{r_c}}\right],\\
\delta(\nu_c)&=\arcsin\!\left[s(\nu_c)\right].
\end{aligned}
\end{equation}
The capped alignment parameter $s$ sets the corresponding angular half-width $\delta$, with the allowed range $|I|<\delta$.

If the signed angle $I$ is taken to be uniform on $[-\pi/2,\pi/2]$, the angular acceptance probability $P_{\rm ang}$ is
\begin{equation}
P_{\rm ang}(|I|<\delta)=\frac{2\delta}{\pi}.
\end{equation}
The corresponding $2\arcsin[\cdots]/\pi$ angular-width form for a randomly oriented circular ring appears in Ref.~\cite{DOr2020Repeated}. The expression corresponds to a uniform distribution of the signed angle $I$, although the reference does not explicitly specify the probability density.

For isotropic orbital normals, $\hat{\mathbf L}_{\rm o}\cdot\hat{\mathbf N}$ is uniform on $[-1,1]$. The relation $\sin[I]=\hat{\mathbf L}_{\rm o}\cdot\hat{\mathbf N}$ gives the angle density $p(I)$ and the accepted fraction $P_{\rm iso}$,
\begin{equation}
p(I)=\frac{1}{2}\cos[I],\qquad
P_{\rm iso}(|I|<\delta)=\sin[\delta]=s.
\end{equation}
The two expressions correspond to different orientation priors. The curves in Fig.~\ref{fig:lensing_probability} retain the uniform signed-$I$ angular-width convention.

A uniform average over $\nu_c$, rather than orbital time, gives
\begin{equation}
\mathcal E(e_{\rm o})=
\frac{1}{2\pi}\int_0^{2\pi}
\sqrt{\frac{1+e_{\rm o}\cos[\nu]}{1-e_{\rm o}^2}}\,\mathrm d\nu.
\end{equation}
Using the identity
\begin{equation}
1+e_{\rm o}\cos[\nu]
=(1+e_{\rm o})\left[1-k^2\sin^2\!\left[\frac{\nu}{2}\right]\right],
\end{equation}
with the elliptic-integral modulus given by
\begin{equation}
k^2=\frac{2e_{\rm o}}{1+e_{\rm o}},
\end{equation}
and the complete elliptic integral of the second kind,
\begin{equation}
E(k)=\int_0^{\pi/2}\sqrt{1-k^2\sin^2[\theta]}\,\mathrm d\theta,
\end{equation}
we obtain
\begin{equation}
\mathcal E(e_{\rm o})=\frac{2E(k)}{\pi\sqrt{1-e_{\rm o}}}.
\label{eq:app_eccentric_factor}
\end{equation}
The expression is exact for the stated uniform average over $\nu_c$, but inserting it into the nonlinear angular-probability formula introduces an approximation. The small-eccentricity expansion is
\begin{equation}
\mathcal E(e_{\rm o})
=1+\frac{7}{16}e_{\rm o}^2+\frac{337}{1024}e_{\rm o}^4
+\mathcal O(e_{\rm o}^6).
\end{equation}
This expansion provides a consistency check of the circular limit. The high-eccentricity curves in Fig.~\ref{fig:lensing_probability} use the full expression in Eq.~\eqref{eq:app_eccentric_factor} rather than the truncated expansion.

The heuristic finite-observation factor adopted from Ref.~\cite{DOr2020Repeated} is
\begin{equation}
\min\!\left[\frac{T_{\rm obs}}{2P_{\rm o}},1\right].
\end{equation}
The resulting estimate is
\begin{equation}
P_{\rm L}\simeq\min\!\left[\frac{T_{\rm obs}}{2P_{\rm o}},1\right]\frac{2}{\pi}\arcsin\!\left[\min\!\left[1,2\eta_{\rm cut}\sqrt{\frac{m_3}{a_{\rm o}}}\mathcal E(e_{\rm o})\right]\right].
\end{equation}
This probability estimate uses the reference Keplerian geometry, the stated averaging prescription, and the adopted orientation prior. The numerical analysis instead follows PN-evolved trajectories and identifies individual lensing events.

\bibliography{references}

@article{Abbott2016GW150914,
  author = {Abbott, B. P. and others},
  collaboration = {LIGO Scientific Collaboration and Virgo Collaboration},
  title = {Observation of Gravitational Waves from a Binary Black Hole Merger},
  journal = {Phys. Rev. Lett.},
  volume = {116},
  pages = {061102},
  year = {2016},
  doi = {10.1103/PhysRevLett.116.061102},
  eprint = {1602.03837},
  archivePrefix = {arXiv}
}

@misc{Colpi2024LISA,
  author = {Colpi, Monica and others},
  collaboration = {LISA Collaboration},
  title = {{LISA} Definition Study Report},
  year = {2024},
  eprint = {2402.07571},
  archivePrefix = {arXiv},
  primaryClass = {astro-ph.CO}
}

@article{Sesana2016Multiband,
  author = {Sesana, Alberto},
  title = {Prospects for Multiband Gravitational-Wave Astronomy after {GW150914}},
  journal = {Phys. Rev. Lett.},
  volume = {116},
  pages = {231102},
  year = {2016},
  doi = {10.1103/PhysRevLett.116.231102},
  eprint = {1602.06951},
  archivePrefix = {arXiv}
}

@article{Toubiana2020StellarLISA,
  author = {Toubiana, Alexandre and Marsat, Sylvain and Babak, Stanislav and Baker, John and Dal Canton, Tito},
  title = {Parameter estimation of stellar-mass black hole binaries with {LISA}},
  journal = {Phys. Rev. D},
  volume = {102},
  pages = {124037},
  year = {2020},
  doi = {10.1103/PhysRevD.102.124037},
  eprint = {2007.08544},
  archivePrefix = {arXiv}
}

@article{Toubiana2021Environmental,
  author = {Toubiana, Alexandre and Sberna, Laura and Caputo, Andrea and Cusin, Giulia and Marsat, Sylvain and Jani, Karan and Babak, Stanislav and Barausse, Enrico and Caprini, Chiara and Pani, Paolo and Sesana, Alberto and Tamanini, Nicola},
  title = {Detectable Environmental Effects in {GW190521}-like Black-Hole Binaries with {LISA}},
  journal = {Phys. Rev. Lett.},
  volume = {126},
  pages = {101105},
  year = {2021},
  doi = {10.1103/PhysRevLett.126.101105},
  eprint = {2010.06056},
  archivePrefix = {arXiv}
}

@article{Inayoshi2017Acceleration,
  author = {Inayoshi, Kohei and Tamanini, Nicola and Caprini, Chiara and Haiman, Zoltan},
  title = {Probing stellar binary black hole formation in galactic nuclei via the imprint of their center of mass acceleration on their gravitational wave signal},
  journal = {Phys. Rev. D},
  volume = {96},
  pages = {063014},
  year = {2017},
  doi = {10.1103/PhysRevD.96.063014},
  eprint = {1702.06529},
  archivePrefix = {arXiv}
}

@article{YuChen2021SMBH,
  author = {Yu, Hang and Chen, Yanbei},
  title = {Direct Determination of Supermassive Black Hole Properties with Gravitational-Wave Radiation from Surrounding Stellar-Mass Black Hole Binaries},
  journal = {Phys. Rev. Lett.},
  volume = {126},
  pages = {021101},
  year = {2021},
  doi = {10.1103/PhysRevLett.126.021101},
  eprint = {2009.02579},
  archivePrefix = {arXiv}
}

@article{Fang2019Spin,
  author = {Fang, Yun and Chen, Xian and Huang, Qing-Guo},
  title = {Impact of a Spinning Supermassive Black Hole on the Orbit and Gravitational Waves of a Nearby Compact Binary},
  journal = {Astrophys. J.},
  volume = {887},
  pages = {210},
  year = {2019},
  doi = {10.3847/1538-4357/ab510e},
  eprint = {1908.01443},
  archivePrefix = {arXiv}
}

@article{KuntzLeyde2023Transverse,
  author = {Kuntz, Adrien and Leyde, Konstantin},
  title = {Transverse Doppler effect and parameter estimation of {LISA} three-body systems},
  journal = {Phys. Rev. D},
  volume = {108},
  pages = {024002},
  year = {2023},
  doi = {10.1103/PhysRevD.108.024002},
  eprint = {2212.09753},
  archivePrefix = {arXiv}
}

@article{Laeuger2024Eccentric,
  author = {Laeuger, Andrew and Seymour, Brian and Chen, Yanbei and Yu, Hang},
  title = {Measuring supermassive black hole properties via gravitational radiation from eccentrically orbiting stellar mass black hole binaries},
  journal = {Phys. Rev. D},
  volume = {109},
  pages = {064086},
  year = {2024},
  doi = {10.1103/PhysRevD.109.064086},
  eprint = {2310.16799},
  archivePrefix = {arXiv}
}

@article{Blanchet2014PN,
  author = {Blanchet, Luc},
  title = {Gravitational Radiation from Post-Newtonian Sources and Inspiralling Compact Binaries},
  journal = {Living Rev. Relativ.},
  volume = {17},
  pages = {2},
  year = {2014},
  doi = {10.12942/lrr-2014-2}
}

@article{ChoPortoYang2022Spin,
  author = {Cho, Gihyuk and Porto, Rafael A. and Yang, Zixin},
  title = {Gravitational radiation from inspiralling compact objects: Spin effects to the fourth post-Newtonian order},
  journal = {Phys. Rev. D},
  volume = {106},
  pages = {L101501},
  year = {2022},
  doi = {10.1103/PhysRevD.106.L101501},
  eprint = {2201.05138},
  archivePrefix = {arXiv}
}

@article{Sberna2022Antichirping,
  author = {Sberna, Laura and Babak, Stanislav and Marsat, Sylvain and Caputo, Andrea and Cusin, Giulia and Toubiana, Alexandre and Barausse, Enrico and Caprini, Chiara and Dal Canton, Tito and Sesana, Alberto and Tamanini, Nicola},
  title = {Observing {GW190521}-like binary black holes and their environment with {LISA}},
  journal = {Phys. Rev. D},
  volume = {106},
  pages = {064056},
  year = {2022},
  doi = {10.1103/PhysRevD.106.064056},
  eprint = {2205.08550},
  archivePrefix = {arXiv}
}

@article{Kidder1995Spin,
  author = {Kidder, Lawrence E.},
  title = {Coalescing binary systems of compact objects to (post)$^{5/2}$-Newtonian order. V. Spin effects},
  journal = {Phys. Rev. D},
  volume = {52},
  pages = {821},
  year = {1995},
  doi = {10.1103/PhysRevD.52.821}
}

@article{BarkerOConnell1975,
  author = {Barker, B. M. and O'Connell, R. F.},
  title = {Gravitational two-body problem with arbitrary masses, spins, and quadrupole moments},
  journal = {Phys. Rev. D},
  volume = {12},
  pages = {329},
  year = {1975},
  doi = {10.1103/PhysRevD.12.329}
}

@article{Apostolatos1994Precession,
  author = {Apostolatos, Theocharis A. and Cutler, Curt and Sussman, Gerald J. and Thorne, Kip S.},
  title = {Spin-induced orbital precession and its modulation of the gravitational waveforms from merging binaries},
  journal = {Phys. Rev. D},
  volume = {49},
  pages = {6274},
  year = {1994},
  doi = {10.1103/PhysRevD.49.6274}
}

@article{BoyerLindquist1967,
  author = {Boyer, Robert H. and Lindquist, Richard W.},
  title = {Maximal Analytic Extension of the {Kerr} Metric},
  journal = {J. Math. Phys.},
  volume = {8},
  pages = {265},
  year = {1967},
  doi = {10.1063/1.1705193}
}

@article{Cutler1998LISA,
  author = {Cutler, Curt},
  title = {Angular resolution of the {LISA} gravitational wave detector},
  journal = {Phys. Rev. D},
  volume = {57},
  pages = {7089},
  year = {1998},
  doi = {10.1103/PhysRevD.57.7089}
}

@article{CornishLarson2001Response,
  author = {Cornish, Neil J. and Larson, Shane L.},
  title = {Space missions to detect the cosmic gravitational-wave background},
  journal = {Class. Quantum Grav.},
  volume = {18},
  pages = {3473},
  year = {2001},
  doi = {10.1088/0264-9381/18/17/308}
}

@article{Liang2019Response,
  author = {Liang, Dicong and Gong, Yungui and Weinstein, Alan J. and Zhang, Chao and Zhang, Chunyu},
  title = {Frequency response of space-based interferometric gravitational-wave detectors},
  journal = {Phys. Rev. D},
  volume = {99},
  pages = {104027},
  year = {2019},
  doi = {10.1103/PhysRevD.99.104027},
  eprint = {1901.09624},
  archivePrefix = {arXiv}
}

@article{Wu2025Kick,
  author = {Wu, Jie and Sun, Mengfei and Ma, Xianghe and Liu, Xiaolin and Li, Jin and Cao, Zhoujian},
  title = {Effect of kick velocity on gravitational wave detection of binary black holes with space- and ground-based detectors},
  journal = {Phys. Rev. D},
  volume = {112},
  pages = {024040},
  year = {2025},
  doi = {10.1103/2qg3-w39y},
  eprint = {2502.13710},
  archivePrefix = {arXiv}
}

@article{Wu2024PNComparison,
  author = {Wu, Jie and Li, Jin and Liu, Xiaolin and Cao, Zhoujian},
  title = {Comparison and application of different post-Newtonian models for inspiralling stellar-mass binary black holes with space-based {GW} detectors},
  journal = {Phys. Rev. D},
  volume = {109},
  pages = {104014},
  year = {2024},
  doi = {10.1103/PhysRevD.109.104014}
}

@article{WuLi2023Confusion,
  author = {Wu, Jie and Li, Jin},
  title = {Subtraction of the confusion foreground and parameter uncertainty of resolvable galactic binaries on the networks of space-based gravitational-wave detectors},
  journal = {Phys. Rev. D},
  volume = {108},
  pages = {124047},
  year = {2023},
  doi = {10.1103/PhysRevD.108.124047}
}

@article{Robson2019LISA,
  author = {Robson, Travis and Cornish, Neil J. and Liu, Chang},
  title = {The construction and use of {LISA} sensitivity curves},
  journal = {Class. Quantum Grav.},
  volume = {36},
  pages = {105011},
  year = {2019},
  doi = {10.1088/1361-6382/ab1101}
}

@article{CutlerFlanagan1994,
  author = {Cutler, Curt and Flanagan, {\'E}anna E.},
  title = {Gravitational waves from merging compact binaries: How accurately can one extract the binary's parameters from the inspiral waveform?},
  journal = {Phys. Rev. D},
  volume = {49},
  pages = {2658},
  year = {1994},
  doi = {10.1103/PhysRevD.49.2658}
}

@article{GergelyKeresztes2003,
  author = {Gergely, L{\'a}szl{\'o} {\'A}. and Keresztes, Zolt{\'a}n},
  title = {Gravitational radiation reaction in compact binary systems: Contribution of the quadrupole-monopole interaction},
  journal = {Phys. Rev. D},
  volume = {67},
  pages = {024020},
  year = {2003},
  doi = {10.1103/PhysRevD.67.024020},
  eprint = {gr-qc/0211027},
  archivePrefix = {arXiv}
}

@article{EggletonKiseleva1995,
  author = {Eggleton, Peter P. and Kiseleva, Ludmila G.},
  title = {An Empirical Condition for Stability of Hierarchical Triple Systems},
  journal = {Astrophys. J.},
  volume = {455},
  pages = {640--645},
  year = {1995},
  doi = {10.1086/176611}
}

@article{Prodan2015Secular,
  author = {Prodan, Snezana and Antonini, Fabio and Perets, Hagai B.},
  title = {Secular Evolution of Binaries near Massive Black Holes: Formation of Compact Binaries, Merger/Collision Products and {G2}-like Objects},
  journal = {Astrophys. J.},
  volume = {799},
  pages = {118},
  year = {2015},
  doi = {10.1088/0004-637X/799/2/118},
  eprint = {1405.6029},
  archivePrefix = {arXiv}
}

@article{Nakamura1998Point,
  author = {Nakamura, Takahiro T.},
  title = {Gravitational Lensing of Gravitational Waves from Inspiraling Binaries by a Point Mass Lens},
  journal = {Phys. Rev. Lett.},
  volume = {80},
  pages = {1138},
  year = {1998},
  doi = {10.1103/PhysRevLett.80.1138}
}

@article{NakamuraDeguchi1999Wave,
  author = {Nakamura, Takahiro T. and Deguchi, Shuji},
  title = {Wave optics in gravitational lensing},
  journal = {Prog. Theor. Phys. Suppl.},
  volume = {133},
  pages = {137--153},
  year = {1999},
  doi = {10.1143/PTPS.133.137}
}

@article{Leung2025WaveOptics,
  author = {Leung, Calvin and Jow, Dylan and Saha, Prasenjit and Dai, Liang and Oguri, Masamune and Koopmans, L{\'e}on V. E.},
  title = {Wave optics, interference, and decoherence in strong gravitational lensing},
  journal = {Space Sci. Rev.},
  volume = {221},
  pages = {29},
  year = {2025},
  doi = {10.1007/s11214-025-01157-7}
}

@article{Takahashi2003Wave,
  author = {Takahashi, Ryuichi and Nakamura, Takashi},
  title = {Wave Effects in the Gravitational Lensing of Gravitational Waves from Chirping Binaries},
  journal = {Astrophys. J.},
  volume = {595},
  pages = {1039--1051},
  year = {2003},
  doi = {10.1086/377430},
  eprint = {astro-ph/0305055},
  archivePrefix = {arXiv}
}

@article{DOr2020Repeated,
  author = {D'Orazio, Daniel J. and Loeb, Abraham},
  title = {Repeated gravitational lensing of gravitational waves in hierarchical black hole triples},
  journal = {Phys. Rev. D},
  volume = {101},
  pages = {083031},
  year = {2020},
  doi = {10.1103/PhysRevD.101.083031},
  eprint = {1910.02966},
  archivePrefix = {arXiv}
}

@article{Yu2021Detecting,
  author = {Yu, Hang and Wang, Yijun and Seymour, Brian and Chen, Yanbei},
  title = {Detecting gravitational lensing in hierarchical triples in galactic nuclei with space-borne gravitational-wave observatories},
  journal = {Phys. Rev. D},
  volume = {104},
  pages = {103011},
  year = {2021},
  doi = {10.1103/PhysRevD.104.103011},
  eprint = {2107.14318},
  archivePrefix = {arXiv}
}

@article{Pijnenburg2024Wave,
  author = {Pijnenburg, Martin and Cusin, Giulia and Pitrou, Cyril and Uzan, Jean-Philippe},
  title = {Wave optics lensing of gravitational waves: Theory and phenomenology of triple systems in the {LISA} band},
  journal = {Phys. Rev. D},
  volume = {110},
  pages = {044054},
  year = {2024},
  doi = {10.1103/PhysRevD.110.044054},
  eprint = {2404.07186},
  archivePrefix = {arXiv}
}

@article{Li2026LT,
  author = {Li, Yu-Zhe and Xu, Wen-Long and Chen, Yi-Gu and Lei, Wei-Hua},
  title = {Lense--Thirring precession modulates repeated lensing of continuous gravitational wave sources in {AGN} disks},
  journal = {Phys. Rev. D},
  volume = {113},
  pages = {043005},
  year = {2026},
  doi = {10.1103/1p77-13n2},
  eprint = {2506.04196},
  archivePrefix = {arXiv}
}

@article{Ubach2025SelfLensing,
  author = {Ubach, Helena and Gieles, Mark and Miralda-Escud{\'e}, Jordi},
  title = {Constraining the environment of compact binary mergers with self-lensing signatures},
  journal = {Phys. Rev. D},
  volume = {112},
  pages = {083026},
  year = {2025},
  doi = {10.1103/ql7q-t6wc}
}

@article{Tamanini2020Acceleration,
  author = {Tamanini, Nicola and Klein, Antoine and Bonvin, Camille and Barausse, Enrico and Caprini, Chiara},
  title = {Peculiar acceleration of stellar-origin black hole binaries: Measurement and biases with {LISA}},
  journal = {Phys. Rev. D},
  volume = {101},
  pages = {063002},
  year = {2020},
  doi = {10.1103/PhysRevD.101.063002},
  eprint = {1907.02018},
  archivePrefix = {arXiv}
}

@article{He2022Moving,
  author = {He, Xiaokai and Liu, Xiaolin and Cao, Zhoujian},
  title = {Gravitational waveform of moving source with high speed},
  journal = {Class. Quantum Grav.},
  volume = {39},
  pages = {195020},
  year = {2022},
  doi = {10.1088/1361-6382/ac8c7e}
}

@article{He2026Accelerating,
  author = {He, Xiaokai and Liu, Xiaolin and Cao, Zhoujian and Chen, Xian},
  title = {Gravitational waveform of constantly accelerating sources},
  journal = {Class. Quantum Grav.},
  volume = {43},
  pages = {105005},
  year = {2026},
  doi = {10.1088/1361-6382/ae66c5}
}

@article{Santos2025BEMRI,
  author = {Santos, Jo{\~a}o S. and Cardoso, Vitor and Nat{\'a}rio, Jos{\'e} and van de Meent, Maarten},
  title = {Gravitational Waves from Binary Extreme Mass Ratio Inspirals: Doppler Shift and Beaming, Resonant Excitation, Helicity Oscillations, and Self-Lensing},
  journal = {Phys. Rev. Lett.},
  volume = {135},
  pages = {211402},
  year = {2025},
  doi = {10.1103/qq3m-6phh}
}

@article{Oancea2024SpinHall,
  author = {Oancea, Marius A. and Stiskalek, Richard and Zumalac{\'a}rregui, Miguel},
  title = {Frequency- and polarization-dependent lensing of gravitational waves in strong gravitational fields},
  journal = {Phys. Rev. D},
  volume = {109},
  pages = {124045},
  year = {2024},
  doi = {10.1103/PhysRevD.109.124045},
  eprint = {2209.06459},
  archivePrefix = {arXiv}
}

@article{Wu2025Globular,
  author = {Wu, Jie and Xiao, Yao and Sun, Mengfei and Li, Jin},
  title = {{Probing Globular Clusters using Gravitational Waves from Inspiraling Stellar-mass Binary Black Holes in Space-based Detector}},
  journal = {Mon. Not. R. Astron. Soc.},
  year = {2026},
  doi = {10.1093/mnras/stag1891},
  eprint = {2508.04021},
  archivePrefix = {arXiv},
  primaryClass = {gr-qc}
}

@article{Wu2026Kerr,
  author = {Wu, Jie and Yao, Jin-Tao and Sun, Mengfei and Li, Jin and Cao, Zhoujian},
  title = {Constraining {Kerr} supermassive black hole properties using gravitational waves from inspiraling stellar-mass binary black holes},
  journal = {Phys. Rev. D},
  volume = {113},
  pages = {124024},
  year = {2026},
  doi = {10.1103/c4jy-7b9k},
  eprint = {2606.12784},
  archivePrefix = {arXiv}
}

@misc{Wu2026AGN,
  author = {Wu, Jie and Sun, Mengfei and Li, Jin and Cao, Zhoujian},
  title = {Constraining {AGN} Disk Properties with Gravitational Waves from Inspiraling Stellar-Mass Binary Black Holes in Hierarchical Triple Systems},
  year = {2026},
  eprint = {2609.00945},
  archivePrefix = {arXiv},
  primaryClass = {gr-qc}
}

@article{Planck2020Cosmology,
  author = {{Planck Collaboration}},
  title = {{Planck} 2018 results. {VI}. Cosmological parameters},
  journal = {Astron. Astrophys.},
  volume = {641},
  pages = {A6},
  year = {2020},
  doi = {10.1051/0004-6361/201833910},
  eprint = {1807.06209},
  archivePrefix = {arXiv}
}

@article{Babak2023Stochastic,
  author = {Babak, Stanislav and Caprini, Chiara and Figueroa, Daniel G. and Karnesis, Nikolaos and Marcoccia, Paolo and Nardini, Germano and Pieroni, Mauro and Ricciardone, Angelo and Sesana, Alberto and Torrado, Jes{\'u}s},
  title = {Stochastic gravitational wave background from stellar origin binary black holes in {LISA}},
  journal = {JCAP},
  volume = {2023},
  number = {08},
  pages = {034},
  year = {2023},
  doi = {10.1088/1475-7516/2023/08/034},
  eprint = {2304.06368},
  archivePrefix = {arXiv},
  primaryClass = {astro-ph.CO}
}

@article{McKechan2010Taper,
  author = {McKechan, D. J. A. and Robinson, C. and Sathyaprakash, B. S.},
  title = {A tapering window for time-domain templates and simulated signals in the detection of gravitational waves from coalescing compact binaries},
  journal = {Class. Quantum Grav.},
  volume = {27},
  pages = {084020},
  year = {2010},
  doi = {10.1088/0264-9381/27/8/084020},
  eprint = {1003.2939},
  archivePrefix = {arXiv}
}

@article{Naoz2016EKL,
  author = {Naoz, Smadar},
  title = {The Eccentric {Kozai--Lidov} Effect and Its Applications},
  journal = {Annu. Rev. Astron. Astrophys.},
  volume = {54},
  pages = {441--489},
  year = {2016},
  doi = {10.1146/annurev-astro-081915-023315},
  eprint = {1601.07175},
  archivePrefix = {arXiv}
}

@article{CutlerVallisneri2007,
  author = {Cutler, Curt and Vallisneri, Michele},
  title = {{LISA} detections of massive black hole inspirals: Parameter extraction errors due to inaccurate template waveforms},
  journal = {Phys. Rev. D},
  volume = {76},
  pages = {104018},
  year = {2007},
  doi = {10.1103/PhysRevD.76.104018}
}

@article{Vallisneri2012RealityCheck,
  author = {Vallisneri, Michele},
  title = {Testing general relativity with gravitational waves: A reality check},
  journal = {Phys. Rev. D},
  volume = {86},
  pages = {082001},
  year = {2012},
  doi = {10.1103/PhysRevD.86.082001}
}

@article{Lindblom2008Accuracy,
  author = {Lindblom, Lee and Owen, Benjamin J. and Brown, Duncan A.},
  title = {Model Waveform Accuracy Standards for Gravitational Wave Data Analysis},
  journal = {Phys. Rev. D},
  volume = {78},
  pages = {124020},
  year = {2008},
  doi = {10.1103/PhysRevD.78.124020}
}

@misc{Huang2026SelfLensing,
  author = {Huang, Jiewei and Ou, Yifan and Hou, Yehui and Guo, Minyong and Chen, Bin},
  title = {Self-lensing of gravitational waves from binary extreme-mass-ratio inspirals: Extended caustics and paired flashes},
  year = {2026},
  eprint = {2610.08399},
  archivePrefix = {arXiv},
  primaryClass = {gr-qc}
}

@article{Buscicchio2025StellarMass,
  author = {Buscicchio, R. and Torrado, J. and Caprini, C. and Nardini, G. and Karnesis, N. and Pieroni, M. and Sesana, A.},
  title = {Stellar-mass black-hole binaries in {LISA}: characteristics and complementarity with current-generation interferometers},
  journal = {JCAP},
  volume = {2025},
  number = {01},
  pages = {084},
  year = {2025},
  doi = {10.1088/1475-7516/2025/01/084},
  eprint = {2410.18171},
  archivePrefix = {arXiv},
  primaryClass = {astro-ph.HE}
}

@article{Cao2014Lensing,
  author = {Cao, Zhoujian and Li, Li-Fang and Wang, Yan},
  title = {Gravitational lensing effects on parameter estimation in gravitational wave detection with advanced detectors},
  journal = {Phys. Rev. D},
  volume = {90},
  pages = {062003},
  year = {2014},
  doi = {10.1103/PhysRevD.90.062003}
}

\end{document}